\documentclass[sigconf]{acmart}
\renewcommand\footnotetextcopyrightpermission[1]{}
\usepackage{multirow}
\usepackage{threeparttable}
\usepackage{color}
\usepackage{soul}
\usepackage{wrapfig}
\usepackage[most]{tcolorbox}  
\usepackage{algorithm}
\usepackage{algpseudocode}
\usepackage{amsmath}
\usepackage{amsfonts}
\usepackage{xcolor}
\usepackage[table]{xcolor}

\newcommand{\name}{COMPASS}
\AtBeginDocument{%
  }

\setcopyright{acmlicensed}
\copyrightyear{2018}
\acmYear{2018}
\acmDOI{XXXXXXX.XXXXXXX}

\acmConference[Conference]{Make sure to enter the correct
  conference title from your rights confirmation email}{June 03--05,
  2018}{Woodstock, NY}

\acmISBN{978-1-4503-XXXX-X/2018/06}

\begin{document}

\title{Bridging Individual and Collective Realism in LLM-Based Human Mobility Simulation via Mobility Scaling-Law Guidance}

\author{Hua Yan}
\affiliation{%
  \institution{Lehigh University}
  \city{Bethlehem}
  \country{USA}}
\email{huy222@lehigh.edu
}
\author{Heng Tan}
\affiliation{%
  \institution{Lehigh University}
  \city{Bethlehem}
  \country{USA}}
\email{het221@lehigh.edu
}

\author{Yu Yang}
\affiliation{%
  \institution{Lehigh University}
  \city{Bethlehem}
  \country{USA}}
\email{yuyang@lehigh.edu
}


\begin{abstract}

Geospatial applications such as urban planning, epidemic forecasting, and transportation demand modeling depend on individual mobility data, but such data are costly to collect, uneven in coverage, and privacy-sensitive. Human mobility simulation offers a scalable alternative. A recent line of work treats large language models (LLMs) as human agents, modeling individual cognitive processes to generate realistic trajectories. Yet because each agent is simulated in isolation, these methods provide no population-level coordination mechanism, and the collective regularities of real mobility — how trip distances, visited locations, and flows distribute across a population — fail to emerge. We close this gap with \name{}, which turns empirical mobility scaling laws into a feedback signal that guides prompt construction. \name{} starts from coarse, population-level adjustments driven by these scaling laws and progressively refines them into individual prompts, jointly satisfying multiple aggregate objectives while keeping individual trajectories realistic. Across two public datasets, \name{} outperforms state-of-the-art LLM-based simulators.

\end{abstract}




\maketitle
\section{Introduction}

Human mobility captures how populations move across geographic space over time, providing essential information for geospatial and spatiotemporal applications in location-based services~\cite{yang2022getnext,yan2023spatio}, traffic forecasting~\cite{liu2023largest,li2017diffusion}, epidemic forecasting~\cite{chang2021supporting,xie2022epignn}, and urban resource allocation~\cite{wang2025coopride}.
In practice, such data come from travel surveys and sensor-based tracking, but both channels have limitations that cap their scalability. Travel surveys record individual trips in detail, yet their high cost and infrequent deployment limit the population they can cover~\cite{toch2019analyzing,bricka2024summary}. Sensor-based tracking, such as mobile-phone traces and Bluetooth beacons, captures finer temporal detail, but its coverage hinges on device adoption and it raises serious privacy concerns~\cite{lajoie2024peoplex}. As a result, neither channel readily delivers mobility data that is both population-scale and privacy-preserving.

Recent studies ~\cite{jiawei2024large,du2025cams,piao2025agentsociety,liu2024human,ju2025trajllm,shao2024chain,bhandari2024urban,li2024geo} prompt large language models (LLMs) to generate synthetic mobility trajectories in a zero-shot way. Unlike approaches that train a trajectory model on real-world data~\cite{feng2020learning, zhu2023difftraj,ouyang2018non}, LLM-based methods enable large-scale data generation while preserving individual privacy.
These methods treat the LLM as a human agent: given a user profile, the model reasons step by step about a person's mobility intentions and produces a trajectory accordingly.
They focus on making each agent’s behavior more human-like by modeling individual-level cognitive processes such as intention and reflection.
These methods, however, simulate each agent on its own, with no mechanism to coordinate behavior across the population. Individual trajectories may therefore look plausible while their aggregate fails to reproduce the mobility patterns seen in real data (Section~\ref{sec:limitation_soa} provides empirical evidence). This gap matters for downstream tasks such as transportation planning and epidemic modeling, where decisions hinge on accurate population-level flows, behavioral heterogeneity, and rare events rather than on individually plausible routines. Our goal is to \textit{develop an LLM-based mobility simulation framework that keeps individual behavior realistic while also matching the population-level patterns found in real data.}

Prior research in social physics shows that human mobility follow stable and reproducible scaling laws at the population level~\cite{brockmann2006scaling,song2010modelling,gonzalez2008understanding, schlapfer2021universal}.
These laws surface in observable measures such as travel distance and radius of gyration, which remain accessible in public mobility datasets even at coarse spatial or temporal resolution.
We argue that these scaling laws of human mobility can serve as population-level guidance for bridging individual-level trajectory generation and collective realism in LLM-based human mobility simulation.
For example, the radius of gyration~\cite{gonzalez2008understanding} quantifies the spatial range of individual mobility, and has been shown to follow a truncated power law.
By comparing the distributions observed in real-world data with those generated by simulation, we can identify differences, such as fewer short-distance trips (around 1 km) and more very long-distance trips (beyond 200 km) in the simulated trajectories.
These differences provide diagnostic signals, indicating which types of individuals are not adequately captured by the generative process.
Based on this analysis, we can tailor individual-level prompts by incorporating explicit behavioral constraints and persona descriptions, so that different individuals are guided by distinct prompts rather than sharing a single prompt that varies only in profile information.
These targeted adjustments correct underrepresented mobility behaviors in the generative process and thereby reproduce the population-level mobility patterns reflected by scaling laws of human mobility.

\begin{figure}[t]
\centering
\includegraphics[width=0.95\linewidth, keepaspectratio=true]{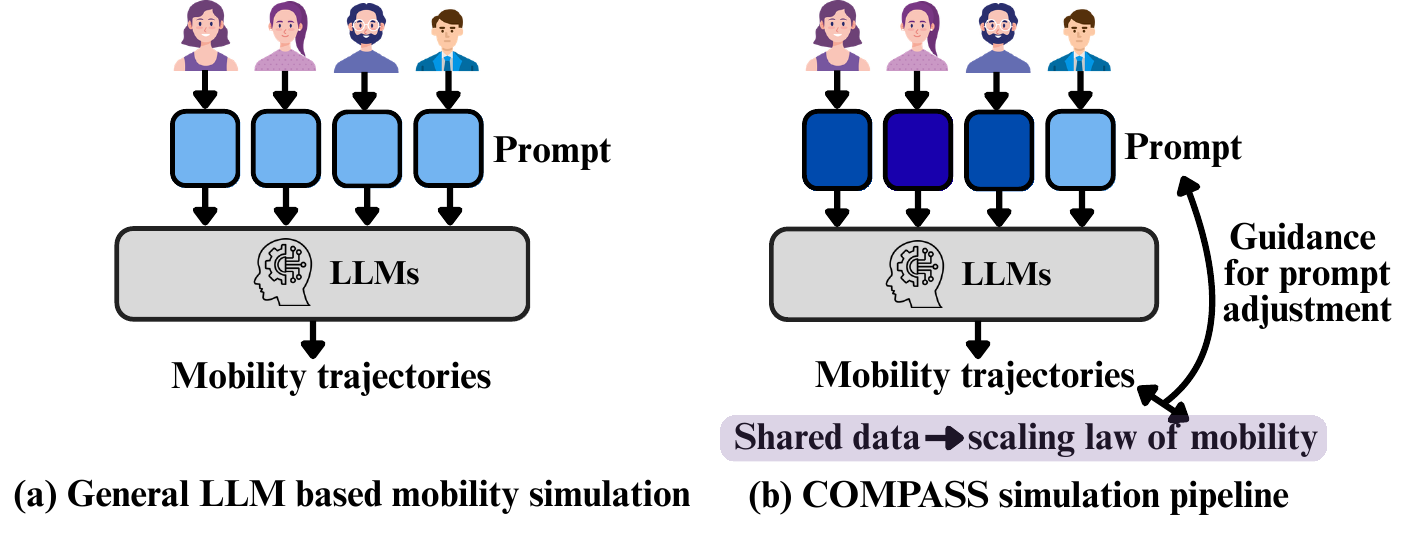}
    % \vspace{-10pt}
    \centering
    \captionsetup{
        font={small},
        skip=2pt  
    }
    \caption{Core idea of \name{}.}
    \label{fig:basic_iead}
\vspace{-25pt}
\end{figure}

Building on this idea, we leverage scaling laws of human mobility to guide individual-level prompt adjustment, generating trajectories that better reproduce population-level patterns, as shown in Figure~\ref{fig:basic_iead}.
Such laws are usually estimated from detailed trajectory data, but we find they survive largely intact in coarse-grained shared datasets (details in Section~\ref{sec:valid_coarse}).
This is consistent with emerging data-sharing practices, where what is released is often coarse trajectories~\cite{yabe2024yjmob100k}, validated simulation output~\cite{yuan2025worldmove}, or aggregated statistics~\cite{gonzalez2008understanding}, and it means our approach can operate under limited data access, improving both generalizability and practical deployability.

Using scaling laws this way is not straightforward. Because they manifest across several facets of population behavior, such as spatial range, temporal regularity, and visitation preference, it is unclear which individuals to adjust and which aspects of their behavior to change. 
Suppose a simulation produces too few short trips and too little activity around midday and at night; this does not mean every adjusted individual should both travel short distances and be active during those hours. Adjustments that satisfy one scaling-law pattern must also be reconciled with the others, since fixing one can easily distort another.

To address these challenges, we view prompt adjustment as a multi-objective, multi-prompt optimization problem, formulated as a Markov Decision Process (MDP) and solved with Monte Carlo Tree Search (MCTS). 
A state is the current set of prompts for all individuals; an action is an LLM-generated coarse-grained adjustment strategy applied to a group of individuals; and the reward measures how closely the resulting trajectories match the scaling-law objectives.
To keep these objectives from competing, we design a multi-objective reward that credits progress on one objective only when it does not degrade the others.
Although each strategy is defined at the group level, we design a two-level MCTS procedure to refine its application at both the strategy and user levels: the search selects promising adjustment strategies while identifying the users most responsive to each strategy. 
Across iterations, individuals accumulate different combinations of adjustments, so coarse group strategies become fine-grained, individual-level adaptations that improve the population-level objectives together.
To reduce cost, we introduce a context-aware global action value estimator to prioritize promising adjustment strategies before expensive trajectory regeneration and evaluation. We further perform the search on a small subset of individuals and then generalize the resulting adjustments to larger groups of individuals with similar profiles.

In particular, our main contributions are as follows.
\begin{itemize}
\item We explore the idea of leveraging scaling laws of human mobility from shared data as guidance to bridge individual trajectory generation and collective realism in LLM-based mobility simulations.

\item We design \name{}, a \underline{CO}llective \underline{M}ulti-\underline{P}rompt \underline{A}djustment framework guided by \underline{S}caling law\underline{S} for large-scale LLM-based human mobility simulation.
Our framework starts from LLM-generated coarse-grained adjustment strategies and uses a two-level MCTS search to decide both which strategies to apply and which individuals should receive each strategy. By allowing different individuals to undergo different combinations of adjustments, \name{} turns coarse strategies into fine-grained individual-level prompt adaptations, which jointly improve multiple scaling-law objectives at the population level.
 
\item  We conduct extensive experiments on two public datasets, and the results demonstrate that our method achieves the best performance, with improvements ranging from 10.70\% to 74.79\% over the best baseline across multiple metrics.
In addition, we evaluate the impact of different types of shared data on our method. The results show that mobility measures obtained from different types of shared data can effectively guide LLM-based mobility simulations.
Notably, even statistical shared data, without trajectory-level information, leads to noticeable performance improvements.
\end{itemize}

\section{Motivation}
\label{sec:motivation}

\subsection{Limitations in reproducing scaling laws}
\label{sec:limitation_soa}
We begin with two classical scaling laws of human mobility, those for travel distance and location visitation frequency, to show that existing simulations cannot fully reproduce the population-level behaviors observed in real data. As a representative case, we compare LLMob~\cite{jiawei2024large}, a recent LLM-based simulator, against real-world data from Beijing (data details in Section~\ref{sec:data}).

\textbf{Travel distance.} Travel distance is the distance $\Delta d$ between consecutive locations in an individual’s trajectory, capturing the spatial scale of human mobility. It follows a truncated power-law,
$P(\Delta d) \sim (\Delta d + \Delta d_0)^{-\beta} \exp\left(-\frac{\Delta d}{\kappa}\right),$
where $\Delta d_0$ is a small offset parameter~\cite{gonzalez2008understanding}.
The exponent $\beta$ determines how frequently long-distance movements occur, with larger values leading to fewer long trips and smaller values leading to more long trips.
The cutoff $\kappa$ captures the spatial extent of individual mobility.

\begin{figure}[h]\centering
\begin{minipage}[b]{0.46\linewidth}
    \includegraphics[width=\linewidth, keepaspectratio=true]{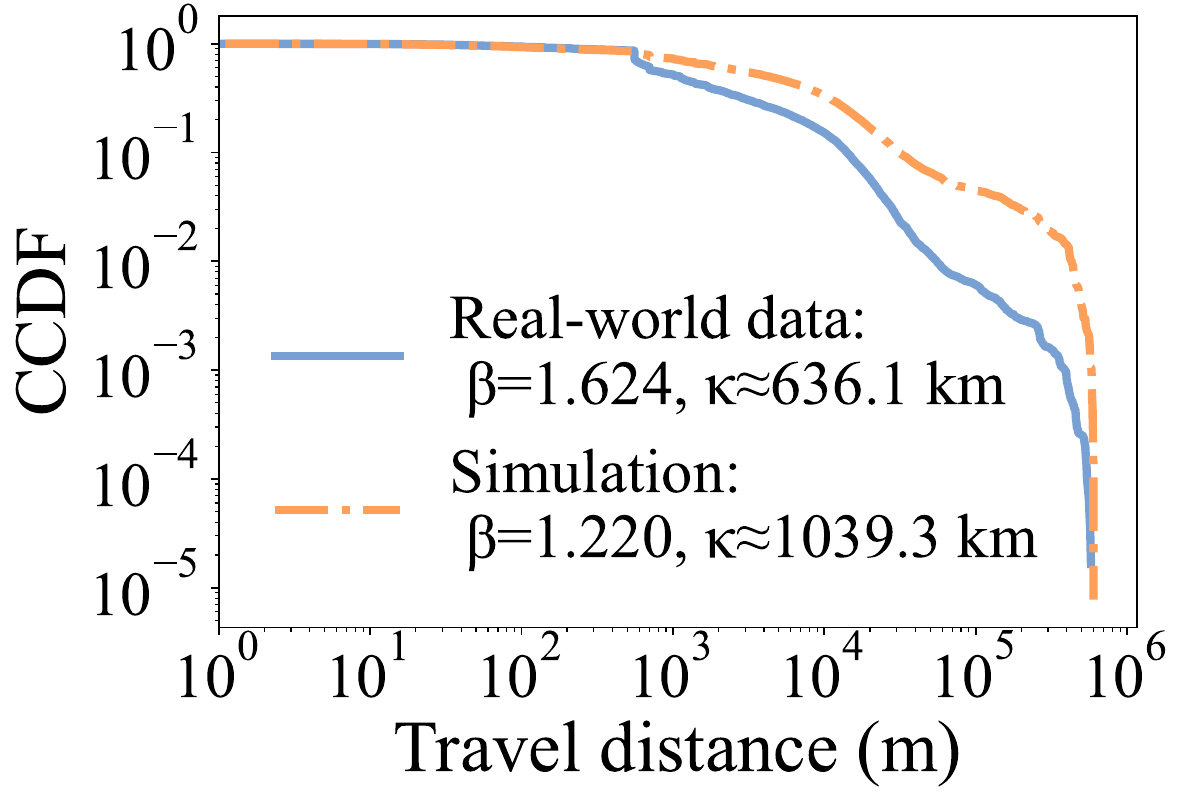}
    \vspace{-10pt}
    \captionsetup{
        font={small},
        skip=2pt  
    }
    \caption{Travel distance distributions (Real vs. Simulation).}
    \label{fig:jump_dismatch}
\end{minipage}
\hspace{10pt}
\begin{minipage}[b]{0.46\linewidth}
    \includegraphics[width=\linewidth, keepaspectratio=true]{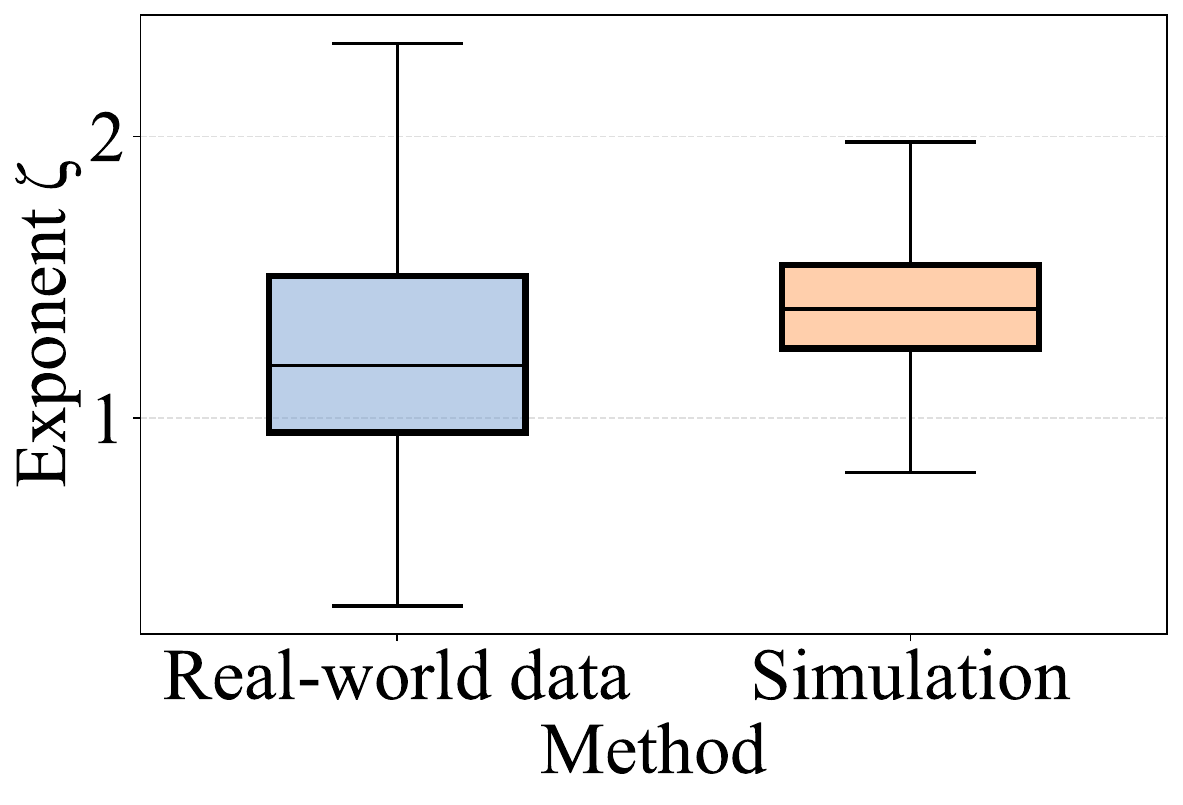}
    \vspace{-10pt}
    \captionsetup{
        font={small},
        skip=2pt  
    }
    \caption{Visitation frequency (Real vs. Simulation).}
    \label{fig:zipf_dismatch}
\end{minipage}
\vspace{-10pt}
\end{figure}

We compute the travel distance from both the real-world data and the simulation and visualize their complementary cumulative distribution functions (CCDFs) in Figure~\ref{fig:jump_dismatch}. 
The real-world data are well described by a truncated power-law, with parameter values consistent with those reported in previous studies~\cite{gonzalez2008understanding} (with $\beta \approx 1.75 \pm 0.15$ and $\kappa \approx 400~\mathrm{km}$). 
In contrast, for the simulation, the estimated scaling exponent is smaller ($\beta \approx 1.22$), and is accompanied by a cutoff at around ($\kappa \approx 1039.3~\mathrm{km}$), which is much larger than that observed in real-world data. 
These results show that the simulation produces too many long-distance trips and does not capture the shape decline in such trips that the real-world data exhibit.

% \vspace{-10pt}

\textbf{Visitation frequency.}
Visitation frequency describes the rank--frequency relationship among the locations an individual visits. 
The visitation frequency $f_k$ of the $k$-th most frequently visited location follows $f_k \propto k^{-\zeta},$
where $\zeta \approx 1.2\pm0.1$ in a prior study~\cite{song2010modelling}. 
A larger value of $\zeta$ indicates that the individual concentrates visits on a small number of core locations, 
whereas a smaller $\zeta$ implies a greater tendency to explore new locations.
For each user, we estimate $\zeta$ and compare the resulting distributions across users between the real-world data and the simulation, as shown in Figure~\ref{fig:zipf_dismatch}.
We observe that the simulation produces insufficient heterogeneity compared with the real-world data. 
In the real-world dataset, some individuals tend to explore a large number of new locations, while others repeatedly return to a small set of core places. 
Together with the travel-distance result, this confirms that the simulation fails to reproduce population-level mobility patterns.

\subsection{A preliminary study on the effectiveness of mobility scaling law guidance}
\begin{figure}[h]\centering
\begin{minipage}[b]{0.46\linewidth}    \includegraphics[width=\linewidth, keepaspectratio=true]{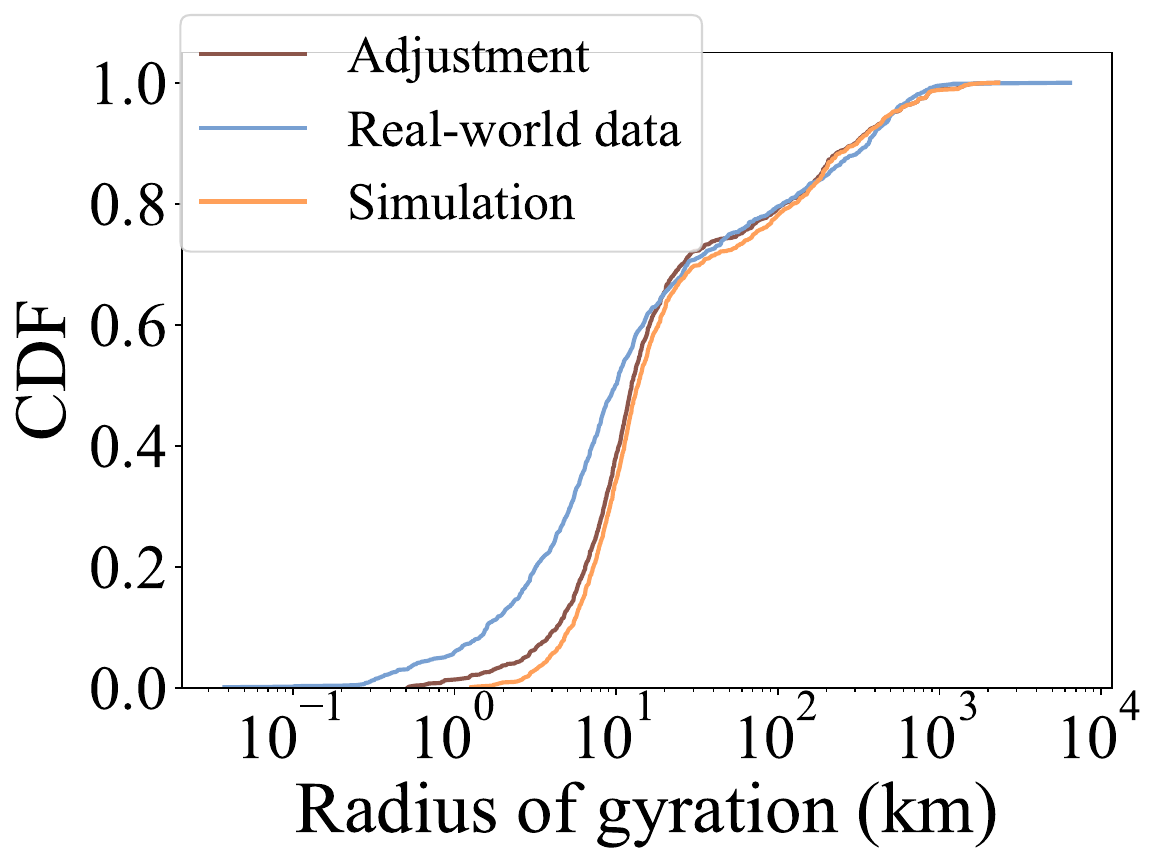}
    \vspace{-15pt}
    \captionsetup{font={small}}
    \caption{Comparison of radius of gyration.}
    \label{fig:radius}
\end{minipage}
\hspace{10pt}
\begin{minipage}[b]{0.46\linewidth}
\includegraphics[width=\linewidth, keepaspectratio=true]{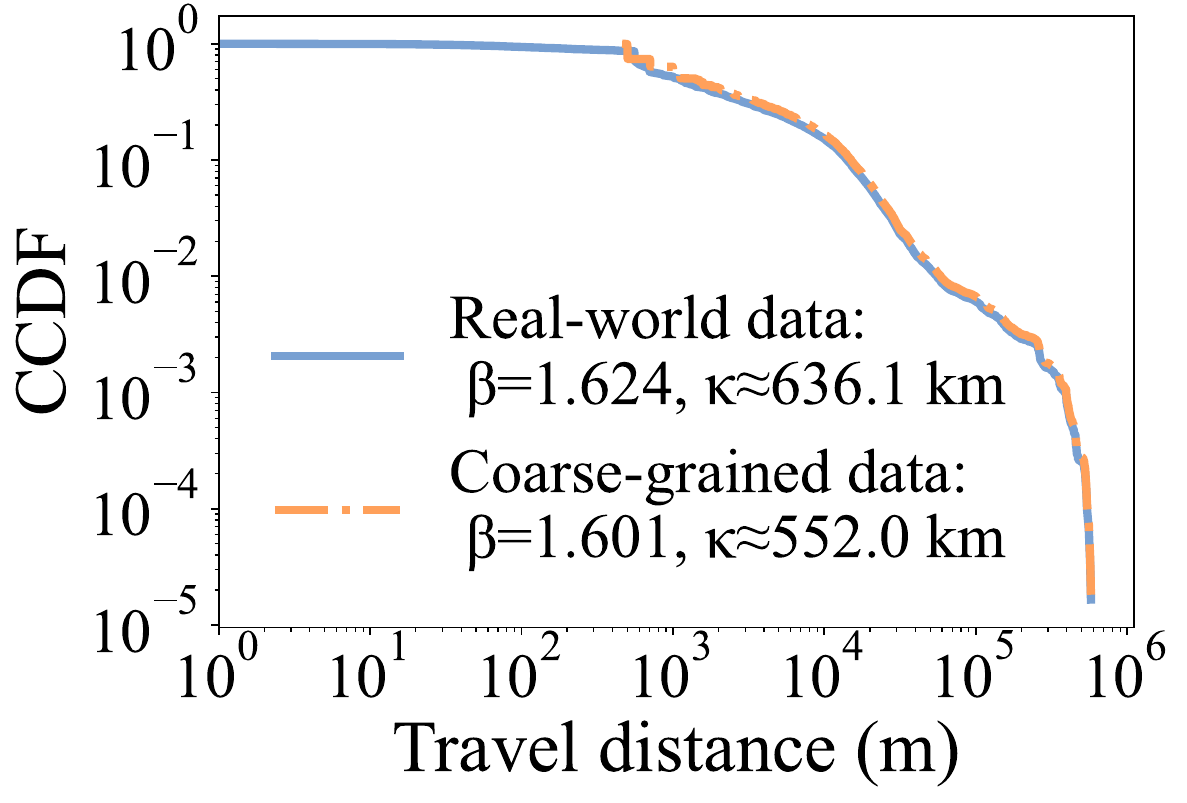}
    \vspace{-15pt}
    \captionsetup{font={small}}
    \caption{Travel distance (Real vs. Coarse-grained).}
    \label{fig:jump_grid}
\end{minipage}
\vspace{-12pt}
\end{figure}

In this section, we provide a preliminary validation of the effectiveness of mobility scaling-law guidance. Specifically, we focus on a single scaling-law objective, the radius of gyration.
We first compute the radius-of-gyration distribution from simulated data. 
Next, we prompt another LLM to analyze the differences between the simulated and target radius distributions. 
Based on this analysis, the LLM provides coarse adjustment suggestions. 
We then randomly select 5\% of individuals for prompt adjustment and generate new trajectories using the updated prompts.
We then visualize the cumulative distribution functions (CDFs) of the radius distributions for the real-world data, the simulation, and the adjusted trajectories, as shown in Figure~\ref{fig:radius}.
We observe that scaling-law-guided adjustments bring the simulated trajectories closer to real-world patterns.
Nevertheless, a discrepancy remains, partly because the adjustment strategy is simple and focuses on only one mobility scaling-law objective. Furthermore, scaling laws of human mobility span multiple dimensions of population-level behavior, making it unclear which specific individuals should be adjusted and whether one or multiple aspects of their behavior need to be modified.

\vspace{-10pt}

\subsection{Preservation of human mobility scaling laws in coarse-grained data}
\label{sec:valid_coarse}
We next ask whether coarse-grained trajectories still carry the scaling laws present in the original real-world data. Following dataset~\cite{yabe2024yjmob100k}, we map the coordinates of the real data onto a $500\,\mathrm{m} \times 500\,\mathrm{m}$ grid and measure all distances in this grid space; time is likewise discretized into intervals of length 48 to form coarse-grained trajectoris.
 On both the real and coarse-grained data, we then compare travel distance, stay duration, and visitation frequency, which together cover the spatial, temporal, and joint spatiotemporal aspects of mobility.
For space, we report only the spatial result here (Figure~\ref{fig:jump_grid}) and defer the rest to Appendix~\ref{sec:supp_motivation}. In every case, the coarse-grained data preserves the same scaling laws as the full-resolution data, and this is what makes coarse and shared data usable as guidance.

\section{Methodology}
\begin{figure*}[t]
    \centering
    \includegraphics[width=\linewidth, keepaspectratio=true]{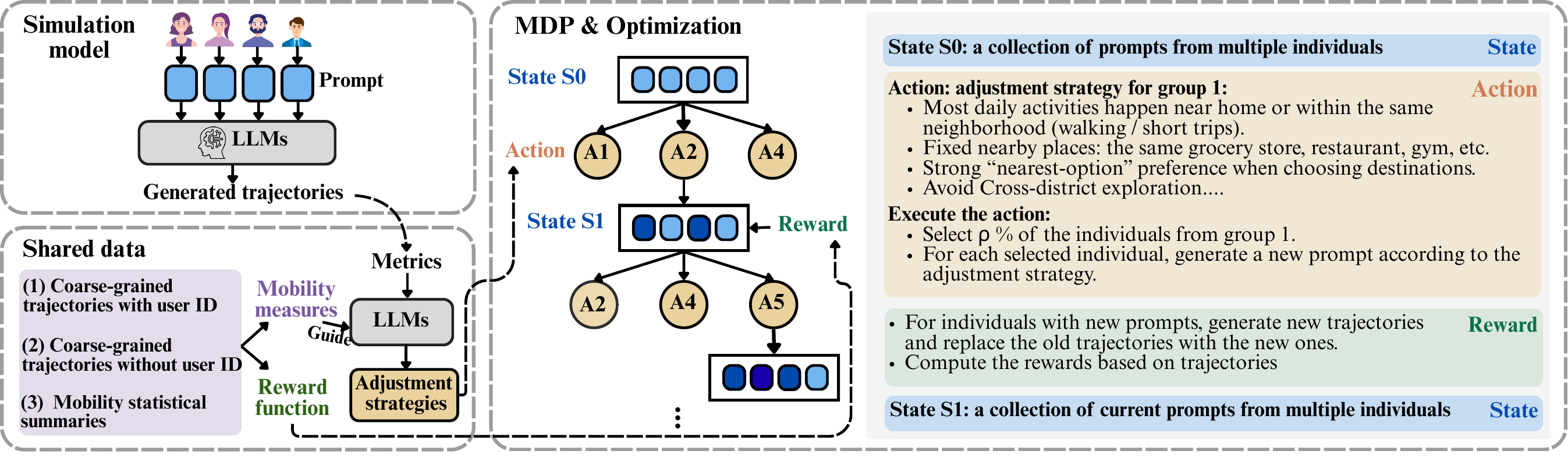}
    \vspace{-25pt}
    \captionsetup{font={small}}
    \caption{Framework of \name.}
    \label{fig:overview}
\vspace{-15pt}
\end{figure*}

\subsection{Definition}
\subsubsection{\textbf{Trajectory.}}
\label{sec:trajectory}

A mobility trajectory refers to a sequence of discrete location visits within a single day, rather than a densely sampled GPS trace. Each visit records where an individual stays at a specific time, marking a meaningful moment in their daily activities. Formally, we define the trajectory of individual $i$ as
$
Y^i = \{ (l_1^i, t_1^i), (l_2^i, t_2^i), \ldots, (l_{T_i}^i, t_{T_i}^i) \},
$
where $l_k^i = (\mathrm{lat}_k^i, \mathrm{lon}_k^i) \in \mathbb{R}^2$ denotes the geographic coordinates and $t_k^i$ denotes the continuous timestamp within the day. The trajectory length $T_i$ is not fixed in advance and varies across individuals, reflecting heterogeneous activity intensities.

\subsubsection{\textbf{LLM-based mobility simulation model}}
\label{sec:sim_model}

We build our simulation model by adapting existing LLM-based mobility simulation approaches~\cite{bhandari2024urban,jiawei2024large}. Following these methods, we model each individual as an LLM-powered agent that generates a mobility trajectory based on a user profile, such as occupation, income, and demographic attributes. The prompt examples are provided in Appendix~\ref{app:prompt}.
Formally, let $\mathcal{M}$ denote the LLM-based simulation model, and let $p_i$ denote the prompt provided to $\mathcal{M}$ for individual $i$. The prompt $p_i$ consists of three components: a user profile and a task description. The simulation process can be expressed as $Y^i = \mathcal{M}(p_i),$
where $Y^i$ is the trajectory generated for individual $i$ as defined in Section~\ref{sec:trajectory}.

Extending this individual-level formulation to population-level simulation, we consider a set of $n$ individuals, each associated with a prompt $p_i$. Let $\mathcal{P} = \{p_1, p_2, \ldots, p_n\}$ denote the set of prompts assigned to these individuals, and let $Y = \{Y^1, Y^2, \ldots, Y^n\}$ denote the resulting trajectories generated by applying $\mathcal{M}$ to each prompt.

\subsubsection{\textbf{Problem formulation}}
\label{sec:problem}

Given the LLM-based simulation model $\mathcal{M}$ and a population of $n$ individuals, whose prompts share the same task description but differ in the user profile, the goal of this work is to identify a set of personalized prompts $\mathcal{P} = \{p^1, \ldots, p^n\}$ that improve the realism of the generated population's mobility behaviors across multiple scaling laws of human mobility, by augmenting each prompt with detailed persona descriptions and behavioral constraints.

Since realism is jointly characterized by multiple scaling law metrics and each individual is assigned its own prompt, we formulate prompt adjustment for human mobility simulation as a \textbf{multi-objective, multi-prompt optimization problem}. Given the scaling law metrics of human mobility from shared data as guidance, we define a set of $D$ target objectives $\mathcal{X}^* = \{x^{*(1)}, \ldots, x^{*(D)}\}$, where each $x^{*(d)}$ corresponds to one such scaling law metric. Our goal is to find an optimal prompt set $\mathcal{P}^*$ that jointly optimizes all objectives in $\mathcal{X}^*$.

\subsection{Overview}
We design \name{}, a multi-prompt adjustment framework for LLM-based human mobility simulation, guided by mobility scaling law from shared data, as shown in Figure~\ref{fig:overview}.
First, we use a simulation model that takes initial prompts as input to generate a set of mobility trajectories.
Second, different types of shared data provide scaling-law guidance for evaluating and improving the generated trajectories (details in Section~\ref{sec:feedback}).
Third, in order to leverage the guidance provided by the shared data to identify an optimal set of prompts for improving simulation, we formulate the prompt adjustment process as a Markov Decision Process, where the state represents the current set of prompts for all individuals, and each action corresponds to a coarse-grained prompt adjustment strategy applied to a group of individuals, generated by an LLM. We then use Monte Carlo Tree Search to find an optimal set of prompts (details in Section~\ref{sec:mdp} and Section~\ref{sec:optimization}). 
Finally, this optimized set of prompts can be used for human mobility simulation and can also be extended to larger populations based on similar profiles, thus reducing the cost of prompt adjustment (details in Section~\ref{sec:usage}).

\subsection{\name{} MDP formulation}
\label{sec:mdp}
\textcolor{blue}{}
We model the prompt adjustment process as a Markov Decision Process (MDP) defined by the tuple $(\mathcal{S}, \mathcal{A}, \mathcal{T}, r)$.
$\mathcal{S}$ denotes the set of states;
$\mathcal{A}$ denotes the action space; 
$\mathcal{T}$ denotes the state transition; 
$r$ is the reward function. 
We introduce $\mathcal{S}, \mathcal{A}, \mathcal{T}, r$ in detail as follows.

\textbf{States $\mathcal{S}$:} The state $s_t \in \mathcal{S}$ represents the current set of prompts assigned to a population of $n$ individuals:$
s_t = \{p_t^1, p_t^2, \dots, p_t^n\}.$
At initialization, all individuals share the same base prompt, i.e., $p_0^i = p_{\mathrm{init}}$, while personalization is achieved by different user profiles.

\textbf{Action $\mathcal{A}$:}
An action $a \in \mathcal{A}$ is a group-specific
prompt-adjustment strategy expressed in natural language,
generated by the procedure described below:

For each shared-data as guidance channel (Sec.~\ref{sec:feedback}), we (i) compute a target mobility scaling law metric (e.g., travel distance) $X^{*}$ from the shared data and the corresponding metric $\hat{X}$ from the current simulated trajectories; 
(ii) prompt an LLM to compare $X^{*}$ with $\hat{X}$, automatically partition the population into $K$ behaviorally coherent groups, and produce one tailored adjustment strategy $a^{(k)}$ for each group. An example prompt is provided in Appendix~\ref{app:prompt}. Each $a^{(k)}$ is treated as a single discrete action (see Figure~\ref{fig:overview} for an example). Repeating the procedure for every target scaling-law metric of human mobility (e.g., spatial and temporal) and aggregating the results yields the final action space $\mathcal{A}$.
The action space is constructed once prior to MCTS and remains fixed throughout the search. Two factors decide $|\mathcal{A}|$: the number of target scaling-law metric of human mobility $D$ (e.g., spatial, temporal) and the number of behavioral groups $K$ the LLM identifies per measure. 
In our experiments this yields $|\mathcal{A}| \approx 10$--$20$.

\textbf{Transition $\mathcal{T}$:}
The transition specifies how the state is updated after applying an action $a_t$. 
Each action targets a specific group of users; we sample a fraction $\rho \in (0, 1]$ of this group via a user-level bandit (details in Sec.~\ref{sec:selection}) to form the selected user set $\mathcal{U}_t$. 
For each $u_i \in \mathcal{U}_t$, the new individual prompt is produced by a prompt-rewriting LLM: $p_{t+1}^i = \mathrm{LLM}_{\text{rewrite}}(p_t^i, a_t)$.
All other individuals keep their previous prompts ($p_{t+1}^i = p_t^i$). The aggregated prompts form the next state $s_{t+1} = (p_{t+1}^1, \dots, p_{t+1}^n)$.

\textbf{Reward $r$:}
The reward function $r$ measures the quality of the simulation at each step.
Let $\mathbf{x}_t = (x_t^{(1)}, \ldots, x_t^{(D)})$ denote the vector of scaling-law metrics of human mobility observed at step $t$, where $x_t^{(i)}$ may be either a scalar or a vector depending on the type of shared data;
let $\mathbf{x}^{*}$ denote the corresponding target vector.
For each scaling-law metric $i$, we define a function $g(x_t^{(i)}, x^{*(i)})$ that measures the distance to the target one, whose specific definition depends on the type of shared data (details in Section~\ref{sec:feedback}).

We normalize each distance function $g(x_t^{(i)}, x^{*(i)})$ by its initial value
$g(x_0^{(i)}, x^{*(i)})$ and aggregate the $D$
per-channel utilities via the geometric mean:
\begin{equation}
  R(t) = \left( \prod_{i=1}^{D}
                \frac{g(x_0^{(i)}, x^{*(i)})}{g(x_0^{(i)}, x^{*(i)}) + g(x_t^{(i)}, x^{*(i)})}
         \right)^{1/D}.
\end{equation}
The geometric mean on bounded utilities enforces balanced
improvements across all dimensions~\cite{jafari2024morl}.
The immediate reward is defined as the increase in aggregated utility between consecutive steps: $r_t = R(t+1) - R(t)$.

\subsection{Scaling-law of human mobility from different types of shared data}
\label{sec:feedback}
We consider three typical types of shared data, each providing different scaling-laws of human mobility for guidance. All such scaling-laws are compatible with our framework; the key difference lies in how they influence action generation and the design of the reward function.

\textbf{(1) Coarse-grained trajectories with user IDs}.
This type of shared data ~\cite{yabe2024yjmob100k} is available at the user level, where each trajectory is linked to a unique but anonymous user ID. 
We select several scaling law metrics of human mobility as the target to guide prompt adjustment. In the following, we describe the selected target scaling law metrics and discuss the reasons for choosing them.
We select the radius of gyration, stay duration, and visitation frequency as target scaling law metrics, representing the spatial, temporal, and spatial-temporal dimensions of the scaling law of human mobility, respectively.
We choose them because they are representative of the three dimensions and have been widely studied in human mobility research, where power-law scaling has often been reported~\cite{gonzalez2008understanding,song2010modelling}.
In addition, these metrics can be computed at the user level, enabling analysis of differences across population groups.

For action generation, we provide the LLM with the CCDF distributions of the radius of gyration and stay duration, as well as the distribution of user-specific $\zeta$ in visitation frequency, for analysis.
For the reward function: we define the function $g(x_t^{(i)}, x^{*(i)})$ measuring the distance between the simulated scaling law metric $x_t^{(i)}$ and its target value $x^{*(i)}$, where $x_t^{(i)}$ is a vector.
This function consists of two components: a scale-sensitive deviation from the target and a distribution discrepancy.
For the first component, we compute the 1-Wasserstein distance in log space, which is defined as
$\mathcal{L}_{\mathrm{W1}}(x_t^{(i)},x^{*(i)})=W_1\bigl(\log(x_t^{(i)}+\varepsilon),\, \log(x^{*(i)}+\varepsilon)\bigr),$
where $\varepsilon$ is a small constant to ensure numerical stability.
For the second component, we measure distance between the two CCDFs: $\mathcal{L}_{\mathrm{L1}}(x_t^{(i)},x^{*(i)}) = \int \bigl|\hat C_t^{(i)}(z) - \hat C_*^{(i)}(z)\bigr| \, \mathrm{d}z,$
where $\hat C_t^{(i)}$ and $\hat C_*^{(i)}$ denote the CCDFs of $x_t^{(i)}$ and $x^{*(i)}$.
Finally, the function $g_i(x_t^{(i)}, x^{*(i)})$ is defined as
$g(x_t^{(i)}, x^{*(i)}) =
\mathcal{L}_{\mathrm{W1}}\!\left(x_t^{(i)}, x^{*(i)}\right)
+ \mu \, \mathcal{L}_{\mathrm{L1}}\!\left(x_t^{(i)}, x^{*(i)}\right)$,
where $\mu$ is a weighting coefficient.

\textbf{(2) Coarse-grained trajectories without user IDs}. 
This type of shared data~\cite{yuan2025worldmove} is not available at the user level and consists only of independent trajectories. Consequently, scaling law metrics that rely on user-level information (e.g, radius of gyration) or long-term mobility history (e.g., revisit probabilities) cannot be computed.
For the target scaling law metrics, we select travel distance and stay duration, which are computed at the trajectory level and represent the spatial and temporal dimensions, respectively.
The travel distance is adopted as the spatial metric, as it reflects individual mobility decisions and directly influences higher-level spatial mobility metrics, such as the radius of gyration and the travel range.

For action generation, we provide the LLM with the CCDF distributions of the travel distance and stay duration, as well as the distribution of user-specific $\zeta$ in visitation frequency, for analysis.
For the reward function, we use the same distance function as in the first type of shared data setting.

\textbf{(3) Mobility statistical summaries}
This type of shared data includes only reported statistical summaries, without explicit trajectory information, e.g., parameters exponent $\beta$ describing travel distance distributions from prior studies~\cite{gonzalez2008understanding}.
Such data provide only coarse-grained guidance for mobility simulations. For example, by comparing the exponent and cutoff parameters of travel distance distributions between real-world and simulated data, we can infer whether certain population groups are potentially overrepresented or underrepresented.
Following the first type of shared data, we select the radius of gyration, stay duration, and visitation frequency as target scaling law metrics.

For action generation, the LLM is provided with statistical summaries, including the exponent and cutoff parameters of the travel distance and stay duration distributions, as well as the total $\zeta$ associated with visitation frequency.
For the reward function: as the scaling law metric $x_t^{(i)}$ is a scalar, we define the function measuring the distance to the target value $x^{*(i)}$ as
$g(x_t^{(i)}, x^{*(i)}) = \frac{\lvert x_t^{(i)} - x^{*(i)} \rvert}{x^{*(i)}}$.
\vspace{-5pt}

\subsection{Optimization}
\label{sec:optimization}
We adapt Monte Carlo Tree Search (MCTS)~\cite{wang2023promptagent} to this MDP framework by incorporating both strategy-level and user-level selection.
In our formulation, each node represents a state and each edge corresponds to a coarse-grained adjustment strategy, while an additional user-level selection step determines which individuals receive the selected strategy. A state-action value function estimates the expected cumulative reward of applying a strategy at a given state. 
The search tree is constructed by iteratively performing four operations: selection, expansion, simulation, and back-propagation.

\textbf{Selection.}
\label{sec:selection}
At each iteration, the algorithm starts from $s_0$ through the search tree until reaching a leaf. Selection operates at two levels: which \emph{action} to take and which \emph{users} the action is applied to.

(1) Action-level:
At a state $s$, among already-expanded actions, we select $a^{*} = \arg\max_{a} \mathrm{UCT}(s, a)$ with the score
$
  \mathrm{UCT}(s, a) = Q(s, a) +
    c \sqrt{\frac{\ln(N(s) + 1)}{N(s, a) + 1}},
$
where $Q(s, a)$ is the mean reward of $(s, a)$, $N(s)$ and $N(s, a)$
are visit counts, and $c$ controls exploration.

(2) User-level:
Given $a^{*}$, the algorithm selects a fraction $\rho$ of users from the group targeted by this action. Rather than sampling randomly, we maintain a per-user reward $Q_{u}(u_i, a, s)$ and visit count $N_{u}(u_i, a, s)$, and select users by
\begin{equation}
  \mathrm{UCB}_{u}(u_i; a^{*}, s) =
    Q_{u}(u_i, a^{*}, s) +
    c_{u} \sqrt{\frac{\ln(N_{a^{*}, s} + 1)}{N_{u}(u_i, a^{*}, s) + 1}},
\end{equation}
where $N_{a^{*}, s} = \sum_j N_{u}(u_j, a^{*}, s)$. This identifies
both the most promising actions and the users most responsive to them.

\textbf{Expansion.}
The expansion step grows the search tree by adding new child nodes under the leaf reached during selection.
Evaluating all actions at every leaf is too costly, since each evaluation triggers multiple LLM calls.
We therefore first filter $\mathcal{A}$ down to a small candidate set $\mathcal{C}$ using a \emph{count-conditioned} global UCB, then evaluate only the candidates in $\mathcal{C}$ by their immediate rewards.

Since repeatedly applying the same action along a path tends to yield diminishing returns, we condition the global statistics on both the action and its in-path occurrence count. 
Let $m_a$ denote the number of times action $a$ has appeared on the path. We keep an average reward $Q_g(a, m_a)$ and visit count $N_g(a, m_a)$, and score each action by its average reward $Q_g(a,m_a)$ only when $N_g(a,m_a)$ reaches a minimum reliability threshold $n_{\min}$; otherwise, we assign it a neutral score of zero.
The top-$n$ actions under this score form the candidate set $\mathcal{C}$.
For each $a \in \mathcal{C}$, the transition function yields a
successor state $s_{a}$ and the reward function yields an immediate
reward $r(s, a)$. We select
$a^{*} = \arg\max_{a \in \mathcal{C}} r(s, a)$, add $s_{a^{*}}$ to
the tree as a new child, and forward it to the simulation stage.

\textbf{Simulation.}
The simulation phase estimates long-term returns by performing a forward rollout starting from the newly expanded state.
Given the initial state $s'$, the algorithm iteratively applies a rollout policy, while no additional tree nodes are created during simulation.
At each rollout step, we first generate a candidate action set $\mathcal{C}$ using the global filtering policy.
For each $a \in \mathcal{C}$, we obtain the resulting state $s_a = \mathcal{T}(s, a)$ and the immediate reward $r(s, a)$ through the transition function and the reward function.
We then select the action with the highest immediate reward to advance the rollout.
The rollout terminates when a predefined stopping criterion is met, such as reaching a maximum rollout depth.

\textbf{Back-propagation.}
After a rollout completes, the observed rewards are propagated
backward along the trajectory to update three sets of value estimates: the per-node $Q(s, a)$ used by UCT, the count-conditioned global $Q_g(a, m_a)$ used by candidate filtering, and the per-user
$Q_{u}(u_i, a, s)$ used by user-level selection.

(1) Per-node update (cumulative).
For each $(s_t, a_t)$ on the trajectory, $Q(s_t, a_t)$ is the average cumulative reward of all $\Omega$ rollouts passing through it: 
$
  Q(s_t, a_t) = \frac{1}{\Omega}\sum_{j=1}^{\Omega}
    \sum_{(s', a') \in \mathrm{traj}_j(s_t, a_t)} r(s', a'),
$
where $\mathrm{traj}_j(s_t, a_t)$ is the sub-trajectory of rollout
$j$ starting at $(s_t, a_t)$ until termination.

(2) Global updates.
We update the global statistics with the immediate reward $r(s_t, a_t)$ instead of the cumulative reward, so that they reflect only the value contributed by $(s_t, a_t)$ itself, without absorbing credit from subsequent actions in the trajectory.

\subsection{Applying the optimized prompt set}
\label{sec:usage}
After identifying the optimal prompts, we can directly use the updated prompt set to generate new mobility trajectories. These prompts can also be scaled to larger populations. 
To reduce the computational cost of searching for the optimal prompt set, we first randomly sample a subset of users and perform prompt optimization on this group. Once the optimized prompts are obtained, we scale them to the full population. 
Specifically, for each adjusted prompt, we assign it to the $m$ most similar users in the full population based on profile similarity, where $m$ is determined by the population size ratio between the subset and the full population.

\section{Evaluation}
\label{sec:eva}

\begin{table*}[t]
\small 
\centering
\renewcommand{\arraystretch}{1}
\setlength{\tabcolsep}{2pt}
\caption{Overall Performance on Beijing dataset. Bold scores are for the best values. The gray-shaded parts indicate the metrics used for guidance under the specific shared data setting, while the remaining metrics are used for evaluation.}
\vspace{-10pt}
\label{tab:overall_bj}
\begin{tabular}{c|ccc|cc|ccc}
\toprule
\multirow{2}{*}{Method} & \multicolumn{3}{c|}{Spatial}                                            & \multicolumn{2}{c|}{Temporal}   & \multicolumn{3}{c}{Spatial-temporal}                                   \\ 
                        & \multicolumn{1}{c}{Radius} & \multicolumn{1}{c}{Distance} & Locfreq & \multicolumn{1}{c}{Duration} & Circadian & \multicolumn{1}{c}{Visit-Freq} & \multicolumn{1}{c}{Exploration} & Return \\ \midrule
CoPB                    & \multicolumn{1}{c}{0.0663$\pm$0.0093}               & \multicolumn{1}{c}{0.0283$\pm$0.0014}       &  0.2548$\pm$0.0630          & \multicolumn{1}{c}{0.0309$\pm$0.0008}       &  0.1229$\pm$0.0336        & \multicolumn{1}{c}{0.1471$\pm$0.0039}            & \multicolumn{1}{c}{0.1445$\pm$0.0275}        & 0.2436$\pm$0.0146      \\ 
UML      & \multicolumn{1}{c}{0.0565$\pm$0.0037}               & \multicolumn{1}{c}{0.0291$\pm$0.0078}       & 0.2711$\pm$0.0173           & \multicolumn{1}{c}{0.0234$\pm$0.0026       }       &  0.0913$\pm$0.0020        & \multicolumn{1}{c}{0.1752$\pm$0.0406}            & \multicolumn{1}{c}{0.1829$\pm$0.0399}        &  0.2669$\pm$0.0735     \\ 
LLMob              & \multicolumn{1}{c}{0.0963$\pm$0.0048}               & \multicolumn{1}{c}{0.0096$\pm$0.0016}       &  0.3111$\pm$0.0083          & \multicolumn{1}{c}{0.0305$\pm$0.0026}       &  0.0906$\pm$0.0005       & \multicolumn{1}{c}{0.1607$\pm$0.0337}            & \multicolumn{1}{c}{0.3895$\pm$0.0290}        &  0.2531$\pm$0.0383     \\ 
CitySim               & \multicolumn{1}{c}{0.0584$\pm$0.0042}               & \multicolumn{1}{c}{0.0531$\pm$0.0044}       &  0.2616$\pm$0.0005          & \multicolumn{1}{c}{0.0325$\pm$0.0018}       &  0.1129$\pm$0.0081        & \multicolumn{1}{c}{0.1507$\pm$0.0181}            & \multicolumn{1}{c}{0.1285$\pm$0.0672}        &   0.3452$\pm$0.0115    \\ \midrule
\name{} w SD2               & \multicolumn{1}{c}{0.0376$\pm$0.0014}              & \multicolumn{1}{c}{\cellcolor{gray!20}\textbf{0.0065$\pm$0.0009}}       &   0.1918$\pm$0.0019        & \multicolumn{1}{c}{\cellcolor{gray!20}0.0178$\pm$0.0003}       &   0.0735$\pm$0.0002       & \multicolumn{1}{c}{0.1315$\pm$0.0020}            & \multicolumn{1}{c}{0.1001$\pm$0.0069}        & 0.0803$\pm$0.0009      \\
\name{} w SD3              & \multicolumn{1}{c}{0.0403$\pm$0.0024}               & \multicolumn{1}{c}{\cellcolor{gray!20}0.0069$\pm$0.0004}       &   0.1957$\pm$0.0058         & \multicolumn{1}{c}{\cellcolor{gray!20}\textbf{0.0174$\pm$0.0005}}       &   0.0732$\pm$0.0003       & \multicolumn{1}{c}{\cellcolor{gray!20}0.1391$\pm$0.0043}            & \multicolumn{1}{c}{0.0854$\pm$0.0018}        &  0.0768$\pm$0.0023     \\\midrule
\name{}               & \multicolumn{1}{c}{\cellcolor{gray!20}\textbf{0.0324$\pm$0.0021}}               & \multicolumn{1}{c}{0.0070$\pm$0.0004}       &  \textbf{0.1914$\pm$0.0016}         & \multicolumn{1}{c}{\cellcolor{gray!20}0.0180$\pm$0.0004}       &  \textbf{0.0718$\pm$0.0007}        & \multicolumn{1}{c}{\cellcolor{gray!20}\textbf{0.1247$\pm$0.0032}}            & \multicolumn{1}{c}{\textbf{0.0756$\pm$0.0134}}        & \textbf{0.0614$\pm$0.0034}      \\ \bottomrule
\end{tabular}
\vspace{-10pt}
\end{table*}

\begin{table}[t]
\footnotesize
\centering
\renewcommand{\arraystretch}{1}
\setlength{\tabcolsep}{3pt}
\caption{Overall Performance on NYC dataset. Bold scores are for the best values. The gray-shaded parts indicate the metrics used for guidance under the specific shared data setting, while the remaining metrics are used for evaluation.}
\vspace{-10pt}
\label{tab:nyc}
\begin{tabular}{c|cc|cc}
\toprule
\multirow{2}{*}{Method} & \multicolumn{2}{c|}{Spatial}                                            & \multicolumn{2}{c}{Temporal}   \\ 
                        & \multicolumn{1}{c}{Distance} & Locfreq & \multicolumn{1}{c}{Duration} & Circadian \\ \midrule
CoPB    & \multicolumn{1}{c}{0.1370$\pm$0.0126}       &    0.2893$\pm$0.0155       & \multicolumn{1}{c}{0.1326$\pm$0.0301}       & 0.1046$\pm$0.0436         \\ 
UML                & \multicolumn{1}{c}{0.1097$\pm$0.0026}       &   0.2812$\pm$0.0339         & \multicolumn{1}{c}{0.1444$\pm$0.0016}       & 0.0666$\pm$0.0069        \\ 
LLMob                      & \multicolumn{1}{c}{0.1395$\pm$0.0048}       &   0.3451$\pm$0.0048         & \multicolumn{1}{c}{0.1400$\pm$0.0005}       &   0.0745$\pm$0.0059       \\ 
CitySim                    & \multicolumn{1}{c}{0.1359$\pm$0.0062}       &   0.2912$\pm$0.0107         & \multicolumn{1}{c}{0.0869$\pm$0.0037}       &  0.1620$\pm$0.0237        \\ \midrule
\name{} w SD2                     & \multicolumn{1}{c}{\cellcolor{gray!20}0.0397$\pm$0.0005}       &  \textbf{0.2511$\pm$0.0503}          & \multicolumn{1}{c}{\cellcolor{gray!20}\textbf{0.0721$\pm$0.0002}}       &   \textbf{{0.0547$\pm$0.0143}}       \\
\name{} w SD3                     & \multicolumn{1}{c}{\cellcolor{gray!20}0.0411$\pm$0.0009}       & 0.2649 $\pm$0.0283         & \multicolumn{1}{c}{\cellcolor{gray!20}0.0854$\pm$0.0005}      & 0.0578 $\pm$ 0.0156        \\ \bottomrule
\end{tabular}
\vspace{-10pt}
\end{table}

\subsection{Dataset description}
\label{sec:data}
We use two public mobility datasets. 
One dataset is collected in Beijing, China, covering the period from October 1, 2019 to December 31, 2019~\cite{shao2024chain}. 
The dataset contains mobility trajectories and user profile information (e.g., age, gender, and occupation), collected via a social networking platform.
The other dataset is a global simulation dataset that has been validated under scaling laws of human mobility~\cite{yuan2025worldmove}. 
All trajectory points are discretized into spatial grids of $1000 \times 1000$ meters, and time is discretized into 48 time slots per day. The dataset does not contain user identifiers. We select one city (New York City) from this dataset, which includes over 100,000 trajectories. Since this dataset does not contain persistent user identifiers, we evaluate only metrics that do not require user-level long-term histories on NYC.
In our simulation setting, user profile data are required as inputs, and mobility trajectories are used for validation.
We select these two datasets because they satisfy both requirements and cover different countries and cities.
Some other publicly available datasets are not suitable for our setting. For example, datasets~\cite{yabe2024yjmob100k} do not release city-level information, making it impossible to construct individual-level profiles. In addition, POI check-in datasets are less appropriate, as their sparse records do not align well with the daily mobility patterns.

For the Beijing dataset, we select 1,200 individuals with approximately one month of data as the ground truth, resulting in around 30,000 trajectories in total. All baseline methods and our approach use the released user profiles and an LLM to generate one-month synthetic human mobility trajectories.
For the New York dataset, we use the entire dataset as the ground truth. We simulate user profiles based on U.S. Census demographics~\cite{census}; all baseline methods and our approach use these simulated profiles and an LLM to generate two-week synthetic human mobility trajectories.

\subsection{Evaluation setup}

\subsubsection{\textbf{Implementation}}
For our method \name{}, we use the first type of shared data as scaling law guidance and implement the framework with Qwen2.5-72B as the LLM backend. 
All experiments are implemented in Python 3.13.5.
For the transition function, we set the user sampling ratio to $\rho=0.15$, meaning that each action is applied to 15\% of users in its corresponding group.
For the reward function, we set $\mu = 1$. 
In the MCTS search, we perform 500 simulations, and the tree depth is determined adaptively by the search process; in practice, the explored trees typically reach a depth of around 8--9. The action-level UCT exploration constant is set to $c=1.4$. For user-level selection, we use a UCB bandit with exploration constant $c_u=1.0$. For candidate filtering, we use count-conditioned global action values with a top-ratio of 0.6, selecting three candidate actions at each non-root expansion.
Following Section~\ref{sec:usage}, for the Beijing dataset, we perform prompt optimization on 30\% of users to reduce search cost, and then extend the optimized prompts to the full population. Specifically, for each optimized prompt, we assign it to the most similar users in the full population based on profile similarity, which is computed using the pre-trained MPNet model~\cite{song2020mpnet}.

\subsubsection{\textbf{Baselines}}
We compare our method with LLM-based human mobility simulation baselines:
\textbf{CoPB~\cite{shao2024chain}} is a mobility simulation framework that guides LLMs through structured reasoning stages to generate realistic mobility intentions.
\textbf{Urban-Mobility-LLM (UML)
~\cite{bhandari2024urban}} is a method that synthesizes travel survey data by prompting LLMs to generate individual mobility patterns.
\textbf{LLMob~\cite{jiawei2024large}} is an LLM-based agent framework for personal mobility generation, combining self-consistency and retrieval strategies.
\textbf{CitySim~\cite{bougie2025citysim}} leverages LLM-powered agents with personas, memory, and long-term goals to simulate realistic human behavior.

We also conduct comparisons with several variants of the model: 
(1) We consider two additional types of shared data that provide scaling law guidance, namely \textbf{\name{} with the second \underline{S}hared \underline{D}ata (w SD2)} and \textbf{\name{} with the third \underline{S}hared \underline{D}ata (w SD3)}. 
(2) \textbf{\name{} w/o PA}. In this setting, we remove the prompt adjustment part.
(3) \textbf{\name{} w E}. In this setting, we perform prompt search on a sampled subset and extend the resulting adjustments to the full population based on profile similarity.
(4) \textbf{\name{} w/o UB}. This variant replaces the per-user UCB bandit with random user sampling when selecting individuals for a chosen adjustment strategy.
(5) \textbf{\name{} w/o UB \& CUCB}. This variant further replaces the count-conditioned global UCB for candidate filtering with a standard action-level UCB that does not distinguish between different application counts of the same adjustment.

\subsubsection{\textbf{Metric}}
Following existing simulation work~\cite{shao2024chain,jiawei2024large}, we evaluate our model using three aspects metrics. 
To assess the similarity between the simulation and real-world data, we compute the Jensen–Shannon divergence (JSD) between their distributions for each metric.
It is worth noting that the evaluation metrics include the scaling law metrics used as guidance. While we report results for all metrics, performance improvements are evaluated only on metrics excluding the guidance metrics.

\textbf{Spatial aspect:} (1) Radius of gyration: quantifies the spatial extent of an individual's mobility. We compute the radius of gyration as the root mean square distance of visited locations from the trajectory centroid.
(2) Travel distance: defined as the geographical distance between consecutive locations in a trajectory.
(3) Origin–destination similarity (OdSim): quantifies the similarity between generated and real trajectories in terms of OD travel patterns. It is computed by comparing the normalized frequency distributions of OD pairs using JSD.

\textbf{Temporal aspect:} 
(1) Stay duration: measures how long individuals remain at each visited location. We compute the duration distribution from the time intervals between consecutive movements.
(2) Circadian rhythm: This metric~\cite{schneider2013unravelling} captures the circadian rhythm of human mobility intensity over a 24-hour period. 
We quantify it by aggregating trips into hourly bins.

\textbf{Spatial-temporal aspect:} 
(1) Visitation frequency.
This metric~\cite{song2010modelling} describes the rank-frequency relationship of an individual’s visited locations.
The detailed definition is in Section~\ref{sec:motivation}.
(2) Exploration:
This metric~\cite{song2010modelling} characterizes how the number of distinct locations grows with the total number of visits. Let $H$ denote the number of stays and $N_{\mathrm{new}}(H)$ the cumulative number of unique locations. Exploration follows a sublinear power-law, $N_{\mathrm{new}}(H) \sim H^{\alpha}$, where $\alpha$ is the exploration exponent. We estimate $\alpha$ for each user and compare its distribution between real and simulated data.
(3) Preferential return: Following~\cite{song2010modelling}, 
we characterize preferential return as the tendency to revisit previously visited locations with probability proportional to their past visitation frequency. 
Let $n_i(t)$ denote the number of visits to location $i$ prior to time $t$. 
Following the original linear assumption ($P \propto n_i$), where $P$ denotes the probability of returning to location $i$, we generalize the model by fitting $P \propto n_i^{\gamma}$. We estimate $\gamma$ for each user and compare the simulated and real user-level $\gamma$ distributions using JSD.

\subsection{Overall performance}
\subsubsection{\textbf{Overall performance on Beijing dataset}}
We compare our method \name{} with other baseline methods, as shown in the Table~\ref{tab:overall_bj}.
For our method \name{}, we use the first type of shared data to provide guidance. 
As guidance, we use three scaling law metrics of human mobility: radius of gyration, stay duration, and visitation frequency, corresponding to the spatial, temporal, and spatio-temporal perspectives, respectively. The guidance metrics are shown in gray, while the remaining metrics are used only for evaluation.
From the Table, we can see that our method outperforms all other baselines, achieving at least an improvement of 27.08\% in travel distance, 24.88\% in odSim, 20.75\% in circadian, 41.17\% in exploration, and 74.79\% in return.
This demonstrates that scaling laws of human mobility can provide effective guidance for individual-level generation, thereby improving collective realism.

\subsubsection{\textbf{Overall performance on NYC dataset}}
For the NYC dataset, we only consider the latter two types of shared data and evaluate spatial and temporal metrics, since the dataset does not provide user-level trajectories and thus does not support spatial-temporal metric computation.
As guidance, we use two scaling law metrics of human mobility: radius of gyration and stay duration, corresponding to the spatial and temporal perspectives, respectively.
The guidance metrics are shown in gray, while the remaining metrics are used only for evaluation. 
We compare our method \name{} with other baseline methods, as shown in Table~\ref{tab:nyc}.
From the Table, we can see that our method outperforms all other baselines, achieving at least an improvement of 10.70\% in OdSim, 17.87\% in Circadian.

\subsection{Performance across shared data types}
In this section, we study how different types of shared data affect prompt adjustment for simulation, and whether good performance can still be achieved using only statistical summaries.

\subsubsection{\textbf{Performance across shared data types on Beijing dataset}}
For the Beijing dataset, we consider three types of shared data and select metrics from three perspectives: spatial, temporal, and spatio-temporal. We compare our method \name{} with variants, as shown in the Table~\ref{tab:overall_bj}.
Due to data type limitations, different shared data settings use different scaling laws of human mobility as guidance. Therefore, we use the metrics not included as guidance for evaluation, as shown in gray.
We can observe that using the first type of shared data  (i.e., \name{}) achieves the best performance, because each scaling law metric is computed at the user level, allowing more precise identification of issues across different individuals. 
The performance of using the third type of shared data (i.e., \name{} w SD3) slightly decreases, since having only mobility statistical summaries can provide only a coarse direction for improvement. Nevertheless, the overall performance of \name{} w SD3 remains good and outperforms all baselines.

\subsubsection{\textbf{Performance across shared data types on NYC dataset}}
For the New York City dataset, we consider two types of shared data and select metrics from spatial and temporal aspects. We compare two variants of our method \name{}, as shown in Table~\ref{tab:nyc}.
Similarly to the Beijing dataset, we use the metrics not included as the guidance for evaluation, as shown in gray.
Overall, using the second shared data (i.e., \name{} w SD2) performs slightly better than using the third shared data (i.e., \name{} w SD3), because \name{} w SD2 provides more fine-grained rewards. Similar to the findings in the Beijing dataset, even when \name{} w SD3 uses mobility statistical summaries as the guidance, performance remains good and outperforms all baselines.

\subsection{Effect of prompt adjustment on mobility scaling laws}
To assess whether scaling-law-guided prompt adjustment improves behavioral realism, we compare simulations before (\name{} w/o PA) and after adjustment (\name{}) on the Beijing dataset under the first shared-data setting, using well-established scaling laws of human mobility: travel distance, preferential return, and exploration scaling.
These analyses complement the metric-based evaluation by examining whether the generated trajectories reproduce key behavioral patterns observed in real-world data.

\begin{figure}[h]
\vspace{-10pt}
    \includegraphics[width=\linewidth, keepaspectratio=true]{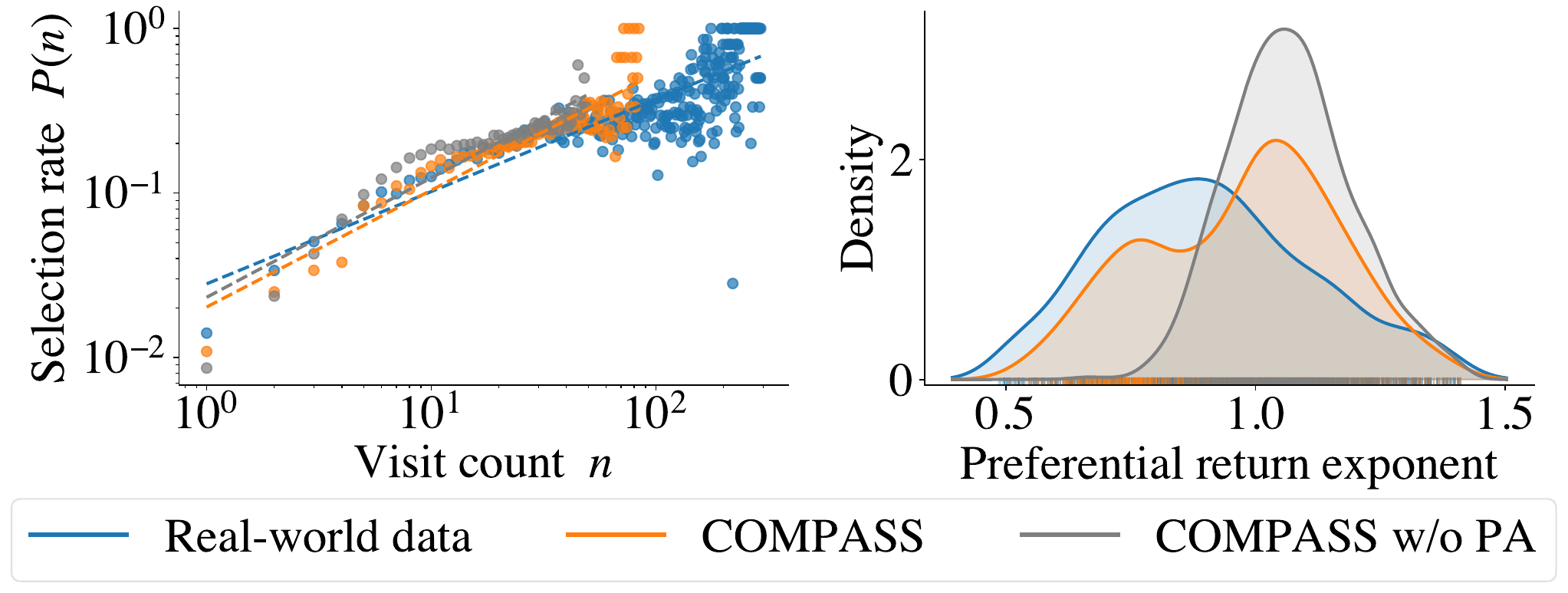}
    \vspace{-20pt}
    \captionsetup{font={small}}
    \caption{Preferential return behavior.}
    \label{fig:preferential_return}
    \vspace{-12pt}
\end{figure}

\textbf{Preferential return:} Figure~\ref{fig:preferential_return} examines preferential return behavior from two perspectives. 
The left panel shows the relationship between a location's visit count $n$ and its selection rate $P(n)$. 
Compared with w/o PA, \name{} more closely matches the real-data trend, particularly for highly visited locations.
The right panel compares the distribution of the preferential-return exponent $\gamma$ estimated separately for each user. 
\name{} produces a distribution that is substantially closer to the real data, while w/o PA shifts toward larger $\gamma$ values. This suggests that our framework not only improves the aggregate
preferential-return pattern but also better captures user-level
heterogeneity in revisit behavior.

\begin{figure}[h]
\vspace{-5pt}
    \includegraphics[width=\linewidth, keepaspectratio=true]{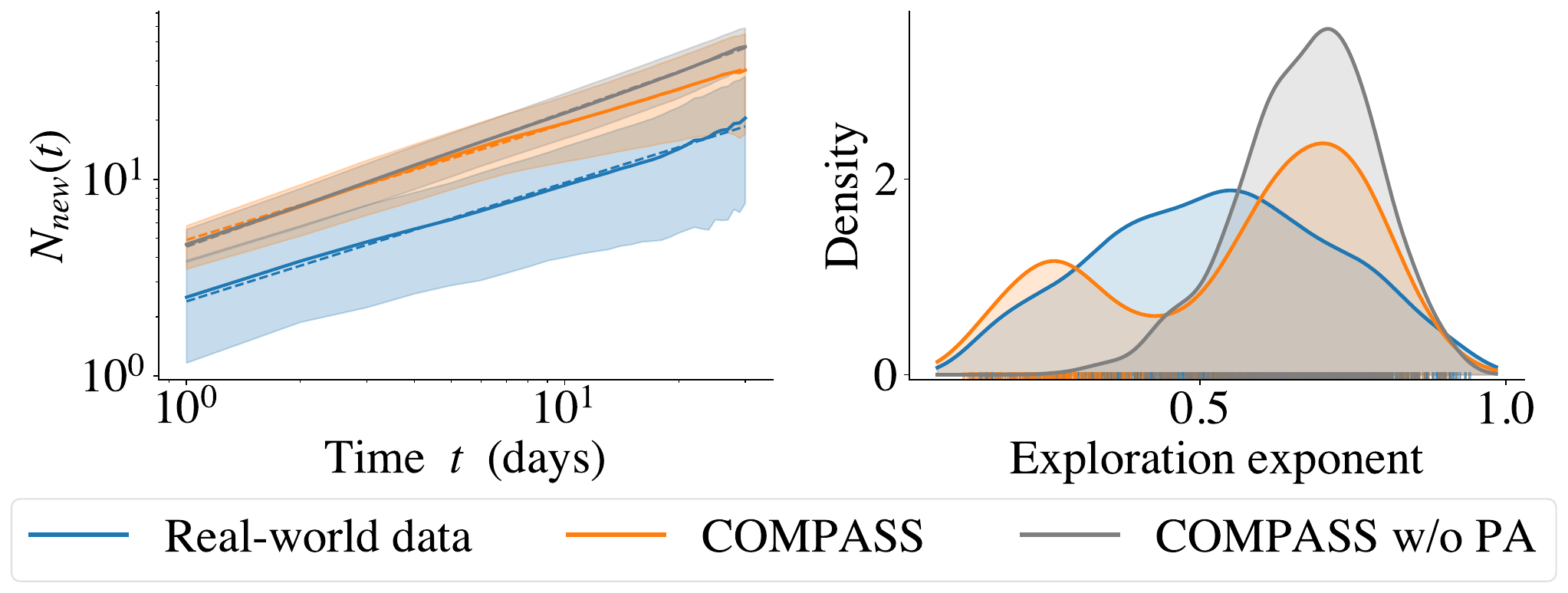}
    \vspace{-20pt}
    \captionsetup{font={small}}
    \caption{Exploration scaling behavior.}
    \label{fig:exploration_scaling}
    \vspace{-12pt}
\end{figure}

\textbf{Exploration:} Figure~\ref{fig:exploration_scaling} examines exploration behavior. The left panel shows the growth of explored locations over time. Although \name{} slightly overestimates exploration compared with the real data, it substantially reduces the gap observed in w/o PA. 
The right panel compares the distribution of the user-level exploration exponent $\alpha$, where \name{} more closely matches the real distribution. These results suggest that our framework improves the modeling of exploration dynamics at both the population and individual levels.

\begin{figure}[h]\centering
\vspace{-10pt}
\begin{minipage}[t]{0.46\linewidth}
    \includegraphics[width=\linewidth, keepaspectratio=true]{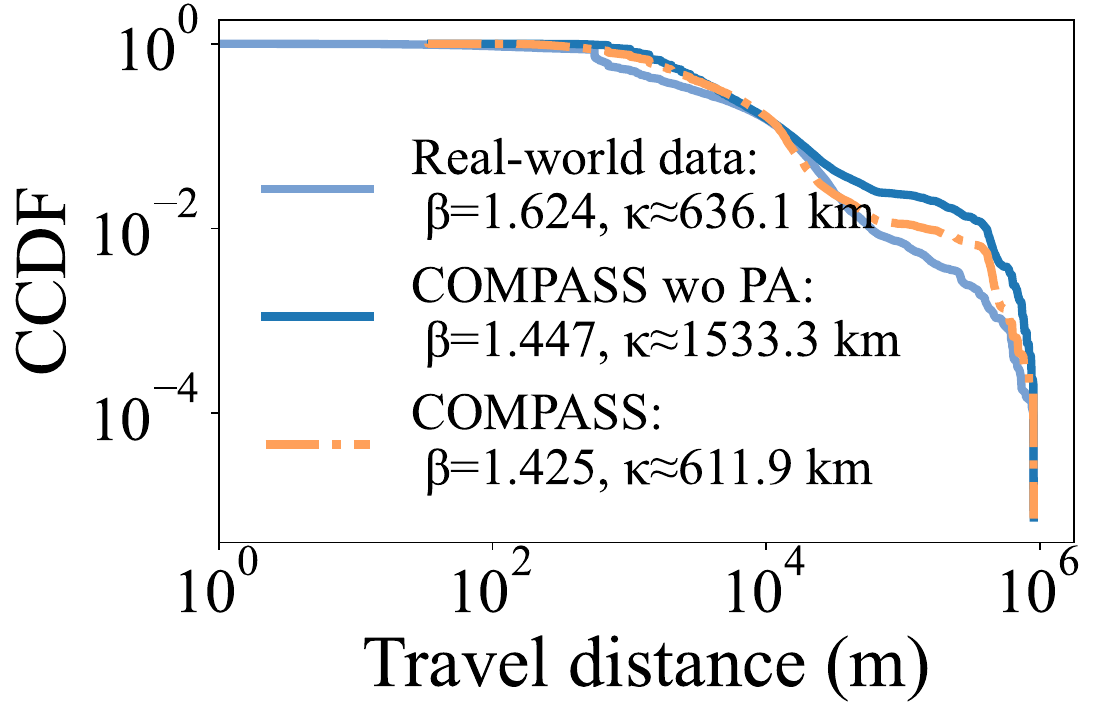}
    \vspace{-15pt}
    \captionsetup{font={small}}
    \caption{Effect of prompt adjustment on travel distance.}
    \label{fig:jump_before_after}
\end{minipage}
\hspace{10pt}
\begin{minipage}[t]{0.46\linewidth}
\includegraphics[width=\linewidth, keepaspectratio=true]{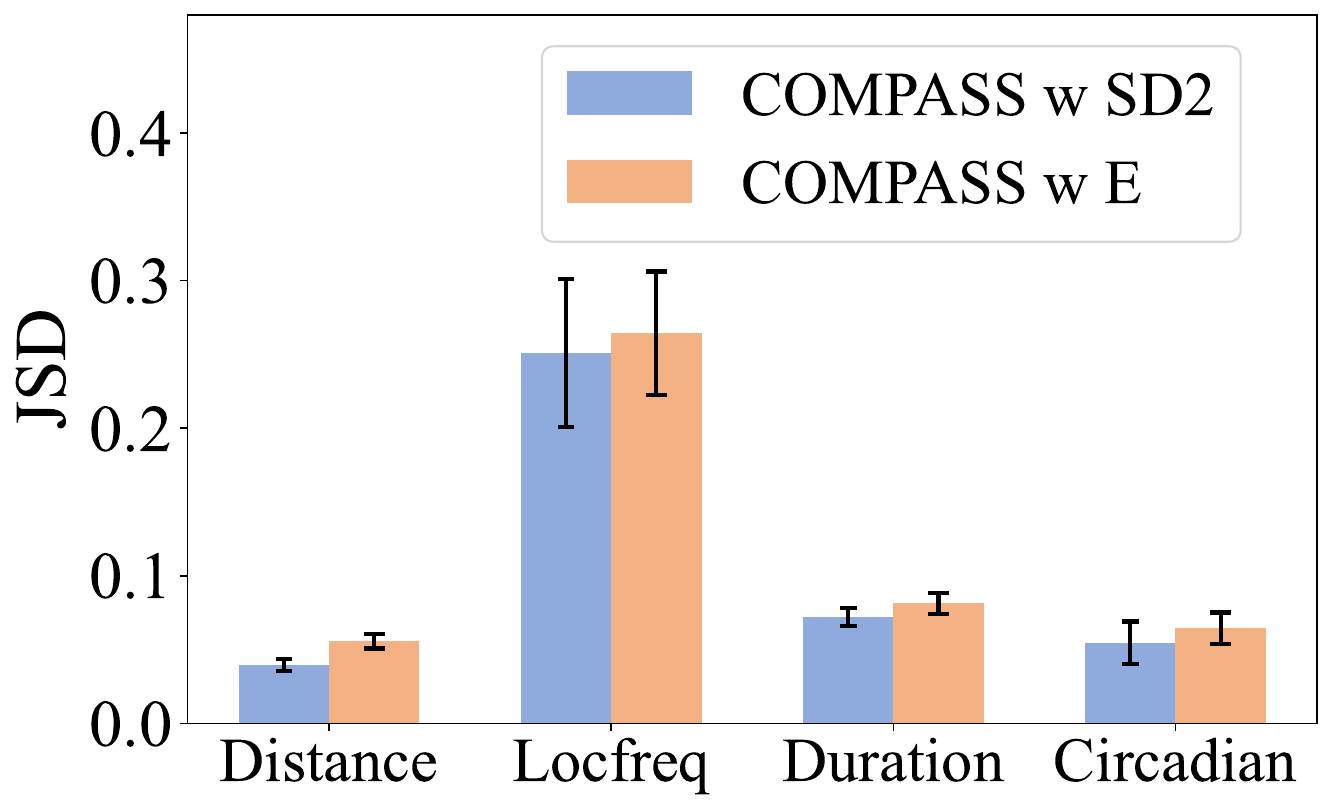}
    \vspace{-15pt}
    \captionsetup{font={small}}
    \caption{Performance of optimal prompt scaling from a subset to the full population.}
    \label{fig:extension}
\end{minipage}
\vspace{-12pt}
\end{figure}

\textbf{Travel distance:} Figure~\ref{fig:jump_before_after} shows the results for travel distance. 
For visualization clarity, travel distances are clipped to a predefined maximum threshold (i.e., 600km) during plotting; all reported metrics are computed using the full dataset.
We observe that, after adjustment, the travel distance distribution becomes closer to the real-world data. 
In addition, the cutoff values obtained from fitting the truncated power-law distribution are more consistent with those of the real-world data, better reflecting realistic human activity ranges.

\subsection{Ablation study}

\subsubsection{Subset-search generalization}
To validate whether the prompt adjustments identified from a small subset can generalize to a larger population, we conduct an experiment using the NYC dataset. 
We choose this dataset because it contains fewer users and trajectories, making full-dataset search computationally feasible. 
Specifically, we compare \name{} w E, which performs prompt search on a subset and extends the resulting adjustments to the full population, with \name{} w SD2, which performs prompt search directly on the full dataset.
Figure~\ref{fig:extension} shows the two variants achieve comparable performance, indicating that prompt adjustments learned from a subset can effectively generalize to larger groups of individuals.

\subsubsection{\textbf{Effect of the user-bandit and count-conditioned UCB}}
To assess the contribution of the two key components in \name{}, we compare the full model with two variants on the Beijing dataset under the first shared-data setting. We report performance on the evaluation metrics that are not directly used as scaling law guidance.
\name{} w/o UB replaces the per-user UCB bandit with random user sampling when selecting users for an action. \name{} w/o UB \& CUCB further replaces the count-conditioned global UCB with a standard action-level UCB that does not distinguish between different application counts of the same adjustment.

\begin{figure}[h]
\vspace{-8pt}
    \includegraphics[width=\linewidth, keepaspectratio=true]{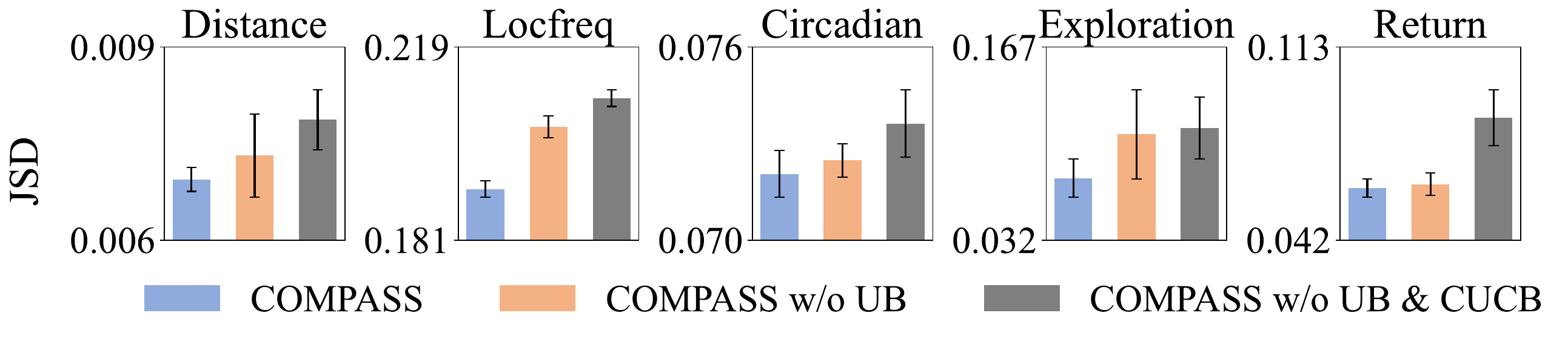}
    \vspace{-20pt}
    \captionsetup{font={small}}
    \caption{Effect of the user-bandit and count-conditioned UCB.}
    \label{fig:ablation:ub&cucb}
    \vspace{-10pt}
\end{figure}

Figure~\ref{fig:ablation:ub&cucb} reports the performance of the full model and its variants on the five evaluation metrics. 
Removing the user-bandit consistently worsens performance across all metrics, while removing both the user-bandit and count-conditioned UCB leads to a further decline. These results confirm that both components contribute to the effectiveness of \name{}.

% \vspace{-5pt}
\subsubsection{\textbf{Effect of MCTS-based optimization}}
\label{sec:search_baselines}
To evaluate the contribution of the proposed MCTS-based search, we compare \name{} with a random-search variant (\name{} w/ RS) on the Beijing dataset under the first shared-data setting. This variant uses the same action space as \name{} but selects actions uniformly at random from the candidate set, without considering future rewards.

\begin{figure}[h]
\vspace{-15pt}
    \includegraphics[width=\linewidth, keepaspectratio=true]{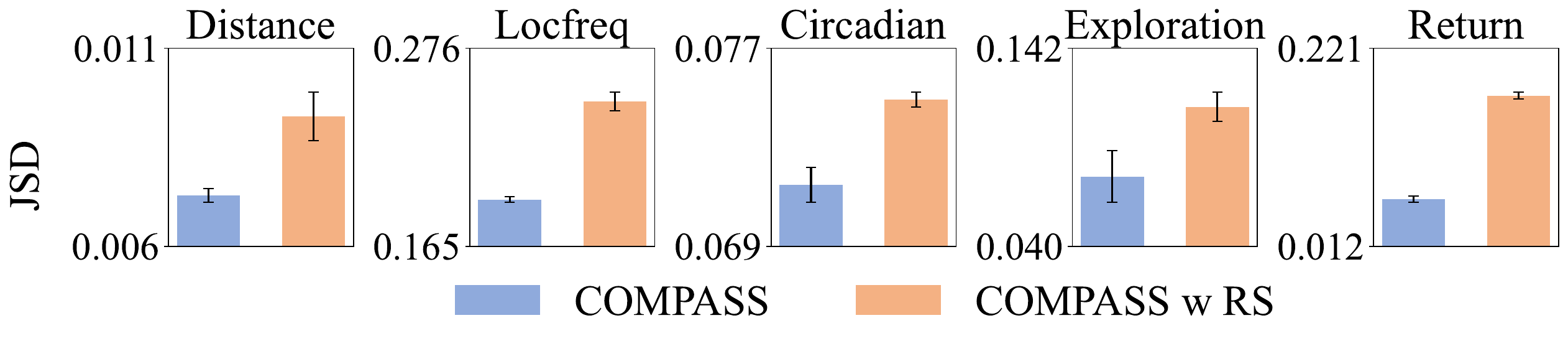}
    \vspace{-20pt}
    \captionsetup{font={small}}
    \caption{Comparison between \name{} and random search.}
    \label{fig:random_search}
    \vspace{-15pt}
\end{figure}

Figure~\ref{fig:random_search} shows that replacing MCTS with random search substantially degrades performance across all metrics. The largest performance drops are observed on Exploration and Return, suggesting that random action selection struggles to capture realistic mobility dynamics.
These results highlight the importance of MCTS-based search. At each optimization step, the framework must determine both what adjustment to apply and which users should receive it. Random search makes these decisions without considering their estimated utility, whereas MCTS explicitly evaluates candidate actions before selecting them. As a result, MCTS is more likely to identify adjustments that improve population-level mobility statistics, leading to consistently lower discrepancies across all metrics.

% \vspace{-5pt} 
\subsection{Performance across different LLMs}
\label{sec:cross_llm}
To examine whether \name{} depends on a particular LLM backbone, we evaluate it on the Beijing dataset under the first shared-data setting using four representative models: GPT-5.2, GPT-4o-mini, LLaMA 3.3 70B, and Qwen 2.5 72B.

Table~\ref{tab:cross_llm} reports the performance of \name{} under four different LLM backbones across mobility metrics. We find that no single model consistently performs best across all metrics. For example, GPT-5.2 achieves the lowest discrepancy on some metrics, while Qwen 72B performs best on others.
More importantly, the differences among the stronger models are relatively small. In particular, Qwen 72B matches GPT-5.2 by achieving the best result on four of the eight metrics. 
This suggests that the effectiveness of \name{} does not rely on a specific LLM backbone. 
Open-source models can serve as a practical alternative to proprietary ones without substantially sacrificing simulation quality.

\begin{table*}[t]
\small 
\centering
\renewcommand{\arraystretch}{1}
\setlength{\tabcolsep}{1pt}
\caption{Performance across different LLMs. Bold scores are for the best values. The gray-shaded parts indicate the metrics used for guidance under the specific shared data setting, while the remaining metrics are used for evaluation.}
\vspace{-10pt}
\label{tab:cross_llm}
\begin{tabular}{c|ccc|cc|ccc}
\toprule
\multirow{2}{*}{Method} & \multicolumn{3}{c|}{Spatial}                                            & \multicolumn{2}{c|}{Temporal}   & \multicolumn{3}{c}{Spatial-temporal}                                   \\ 
                        & \multicolumn{1}{c}{Radius} &  Distance &\multicolumn{1}{c|}{Locfreq} & \multicolumn{1}{c}{Duration} & Circadian & \multicolumn{1}{c}{Visit-Freq} & \multicolumn{1}{c}{Exploration} & Return \\ \midrule

\name{} w GPT5.2               & \multicolumn{1}{c}{\cellcolor{gray!20}0.0385$\pm$0.0026}         &\textbf{0.0064$\pm$0.0005}     & \multicolumn{1}{c|}{0.2017$\pm$0.0076}               & \multicolumn{1}{c}{\cellcolor{gray!20}0.0221$\pm$0.0002}       &   0.0738$\pm$0.0003       & \multicolumn{1}{c}{\cellcolor{gray!20}\textbf{0.1095$\pm$0.0042}}            & \multicolumn{1}{c}{\textbf{0.0645$\pm$0.0014}}        & \textbf{0.0554$\pm$0.0101}      \\
\name{} w LLama3.3              & \multicolumn{1}{c}{\cellcolor{gray!20}0.0403$\pm$0.0008}           &0.0068$\pm$0.0001    & \multicolumn{1}{c|}{0.2081$\pm$0.0018}               & \multicolumn{1}{c}{\cellcolor{gray!20}0.0223$\pm$0.0002}       &   0.0731$\pm$0.0002      & \multicolumn{1}{c}{\cellcolor{gray!20}0.1746$\pm$0.0093}            & \multicolumn{1}{c}{0.0618$\pm$0.0030}        &  0.0745$\pm$0.0018     \\

\name{} w GPT4o-mini              & \multicolumn{1}{c}{\cellcolor{gray!20}0.0410$\pm$0.0435}          &0.0077$\pm$0.0010     & \multicolumn{1}{c|}{0.2274$\pm$0.0106}               & \multicolumn{1}{c}{\cellcolor{gray!20}0.0229$\pm$0.0011}       &   0.0721$\pm$0.0007       & \multicolumn{1}{c}{\cellcolor{gray!20}0.1612$\pm$0.0072}            & \multicolumn{1}{c}{0.0751$\pm$0.0239}        &  0.1129$\pm$0.0606     \\

\name{} w Qwen               & \multicolumn{1}{c}{\textbf{\cellcolor{gray!20}0.0324$\pm$0.0011}}              &0.0070$\pm$0.0002 & \multicolumn{1}{c|}{\textbf{0.1914$\pm$0.0016}}       & \multicolumn{1}{c}{\cellcolor{gray!20}\textbf{0.0180$\pm$0.0004}}       &  \textbf{0.0718$\pm$0.0007}        & \multicolumn{1}{c}{\cellcolor{gray!20}0.1247$\pm$0.0032}            & \multicolumn{1}{c}{0.0756$\pm$0.0134}        & 0.0614$\pm$0.0034     \\ \bottomrule
\end{tabular}
\vspace{-10pt}
\end{table*}

\vspace{-5pt}
\subsection{Parameter sensitivity analysis}
\label{sec:sensitivity}
\begin{figure}[h]
\vspace{-10pt}
    \includegraphics[width=\linewidth, keepaspectratio=true]{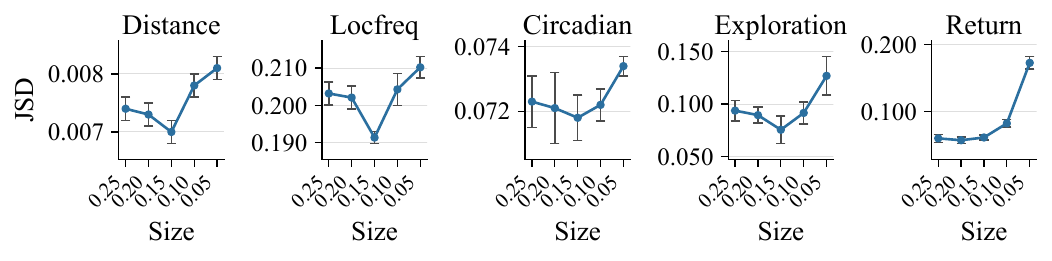}
    \vspace{-20pt}
    \captionsetup{font={small}}
    \caption{Sensitivity of \name{} to the sampling rate.}
    \label{fig:sensitivity}
    \vspace{-15pt}
\end{figure}

The sampling rate $k$ controls the fraction of users selected for each action. A smaller $k$ reduces the number of LLM calls but provides less feedback for evaluating candidate adjustments, while a larger $k$ increases computational cost by applying adjustments to more users.
We vary $k \in \{0.05,0.10,0.15,0.20,0.25\}$ on the Beijing dataset under the first shared-data setting and report the performance on its five corresponding evaluation metrics in Figure~\ref{fig:sensitivity}.

Figure~\ref{fig:sensitivity} shows that $k=0.15$ achieves the best overall performance, yielding the best results on four of the five evaluation metrics.
Both smaller and larger values of $k$ lead to worse results, suggesting that an intermediate sampling rate provides the best balance between evaluation quality and adjustment granularity.

\vspace{-5pt}
\subsection{Convergence analysis}
\label{sec:convergence}
\begin{figure}[h]
\vspace{-10pt}
        \includegraphics[width=0.8\linewidth, keepaspectratio=true]{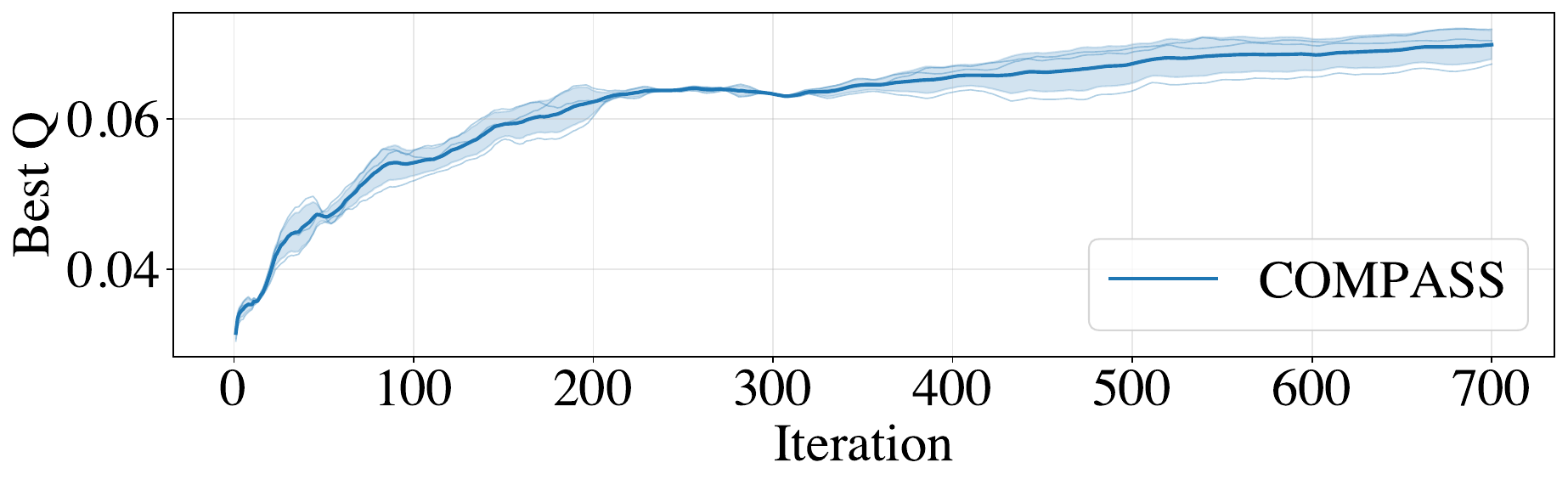}
    \vspace{-15pt}
    \captionsetup{font={small}}
    \caption{MCTS convergence on the Beijing dataset.}
    \label{fig:convergence}
    \vspace{-10pt}
\end{figure}

To examine whether the MCTS optimization converges within a fixed number of iterations, we track the best Q value during the search on the Beijing dataset under the first shared-data setting. Figure~\ref{fig:convergence} shows the rolling mean ($\pm$ std), computed using a 10-iteration window.
The \name{} curves begin to plateau after roughly 500 iterations, with only marginal improvements thereafter. Therefore, we set 500 iterations as the default optimization budget for all experiments.

\subsection{Overhead analysis}
We quantify the overhead using two metrics: the number of LLM calls and the monetary cost. In our implementation, one LLM call refers to either one prompt-adjustment call that updates a user’s behavioral traits, or one trajectory-generation call that generates a one-day trajectory for a user. Thus, the number of LLM calls measures the amount of fresh generation required during search, while the cost is estimated based on the provider’s token pricing.

To reduce repeated LLM calls, we use a user-level cache. Each time an action is executed, we cache the before-and-after adjustment state of the selected users together with their regenerated trajectories. If the same action is later applied under the same state and selects a user already stored in the cache, we directly reuse the cached trajectory instead of invoking the LLM.
In 500 MCTS iterations, the cache achieved a 95.9\% hit rate, reducing the search overhead to an average of 31.5 LLM calls per iteration. Under Qwen2.5 pricing, this corresponds to a monetary cost of \$5.97.

\section{Discussion}
\label{sec:Dis}
\textbf{Lessons learned:}
Based on the results of our paper, we summarize the following lessons learned:
(1) Scaling laws of human mobility obtained from different types of shared data can effectively guide LLM-based mobility simulations, as shown in Table~\ref{tab:overall_bj} and Table~\ref{tab:nyc}. Even statistical information without any trajectories leads to noticeable performance improvements.
(2) Our framework incorporates multi-dimensional scaling law metrics as guidance, enabling improvements that extend beyond targeted aspects and enhance the overall mobility simulation, as shown in Table~\ref{tab:overall_bj}.
(3) Our framework enables identifying optimal prompt combinations from a small subset of individuals and generalizing them to a larger population, reducing search overhead, as shown in Figure~\ref{fig:extension}.

\textbf{Limitation:} Despite its effectiveness, our framework has several limitations. 
Our framework introduces additional computational overhead in mobility simulation, particularly at large scales, although this cost is partially mitigated by our method.
Our evaluation is currently limited to two cities and extending validation to more diverse urban environments remains future work.

\textbf{Ethics and privacy:} 
This work focuses on population-level mobility simulation rather than individual tracking or prediction. Our framework uses user profiles as inputs, which are derived from census statistics or publicly available data. All information is highly anonymized and represented at a coarse-grained level, preventing identification of individuals. Scaling laws of human mobility are obtained from publicly shared data and do not involve privacy-sensitive information. While LLM-based simulations may introduce biases or unrealistic behaviors, our framework uses scaling laws of mobility to guide individual generation for improving realism.

\section{Related work}

\textbf{LLM-based human mobility simulation.}
Existing works explored using LLMs for human mobility simulation by leveraging their knowledge and human-like reasoning capabilities. The advantage of these approaches is that they do not require large-scale real-world mobility data for training.
These works~\cite{jiawei2024large,du2025cams,piao2025agentsociety,mou2024individual,liu2024human,ju2025trajllm,shao2024chain,bhandari2024urban,li2024geo} guide LLMs to simulate human-like mobility intention reasoning step by step and then produce realistic mobility activity sequences. 
For example, CoPB~\cite{shao2024chain} is an intention- and planning-based framework that enables LLMs to generate human mobility trajectories through step-by-step reasoning. 
Although these works produce realistic outputs, the reliance on LLMs makes the simulation expensive.
This work~\cite{jiawei2024large} designs a LLM-based agent framework for personal mobility generation, combining self-consistency and retrieval strategies to align language models with real-world human activity.
However, they generate each individual's trajectory independently, without any population-level coordination mechanism, thereby failing to capture the emergence of collective behaviors.

\textbf{Prompt optimization}
Previous research on prompt optimization focuses on identifying an optimal prompt that enables an LLM to produce the best performance for a given class of tasks~\cite{wu2024strago,cheng2024black,trivedi2025align}. Existing methods can be categorized into three types.
The first category focuses on how to select the optimal prompt given available candidates~\cite{hu2024localized,shi2024efficient}. For example, given a set of human-readable prompt candidates with unknown performance, ZOPO~\cite{hu2024localized} embeds prompts into a continuous space and estimates update directions based on past performance, enabling efficient local exploration toward high-performing prompts.
The second category considers both prompt refinement and optimal prompt selection, and addresses single-objective optimization~\cite{wang2023promptagent,yang2024ampo}. For example, Promptagent~\cite{wang2023promptagent} enables an LLM to iteratively refine prompts based on observed errors, using MCTS to explore the space of prompt modifications.
The third category considers both prompt refinement and prompt selection, with a focus on multi-objective optimization~\cite{zhao2025pareto,jafari2024morl,singla2024dynamic,yuan2024pre,sinha2024survival}. 
These works focus on balancing multiple objectives. For example, ParetoPrompt~\cite{zhao2025pareto} uses a pre-trained language model as a policy, which is further optimized via reinforcement learning to achieve multi-objective prompt optimization. 
Unlike prior work that focuses on finding a single prompt to achieve good performance on a specific task (single or multiple objectives), we aim to address a multi-prompt, multi-objective optimization problem. Specifically, our goal is to find an optimal set of prompts that jointly optimizes multiple objectives.
\section{Conclusion}
In this work, we design \name{} for LLM-based mobility simulation. It leverages scaling laws of human mobility from shared data as guidance to adjust individual-level prompts in order to reproduce realistic population-level mobility behavior. 
Our framework applies coarse-grained adjustment strategies guided by scaling laws of human mobility, progressively enabling fine-grained individual-level adaptation while satisfying multiple population-level mobility objectives under a limited budget.
Experiments show that \name{} significantly outperforms other LLM-based simulations.

\newpage
\bibliographystyle{ACM-Reference-Format}
\bibliography{sample-base}
\newpage

\appendix
\section{Appendix}

\subsection{Supplementary analysis: preservation of human mobility scaling laws in coarse-grained data}
\label{sec:supp_motivation}
\begin{figure}[h]\centering
\begin{minipage}[b]{0.48\linewidth}
    \includegraphics[width=\linewidth, keepaspectratio=true]{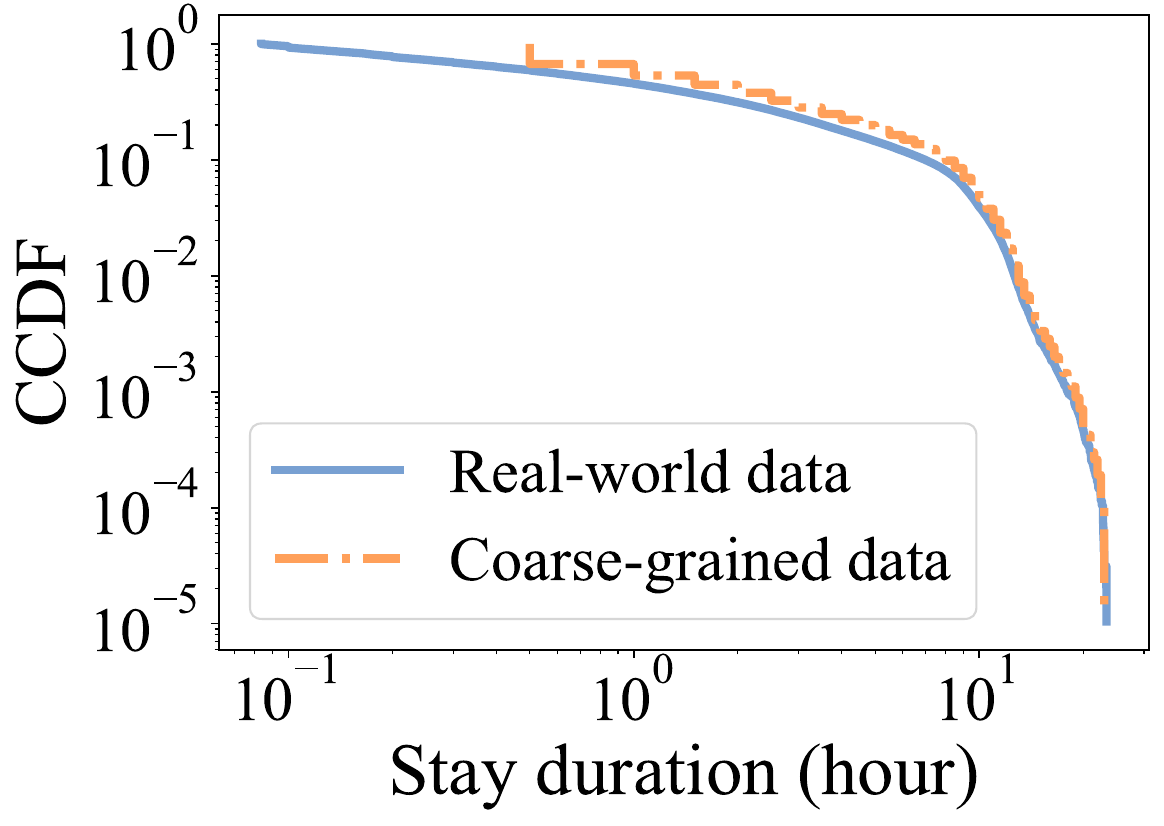}
    \vspace{-15pt}
    \captionsetup{font={small}}
    \caption{\small Stay duration distributions (Real vs. Coarse-grained).}
    \label{fig:duration_grid}
\end{minipage}\hfill
\begin{minipage}[b]{0.48\linewidth}
    \includegraphics[width=\linewidth, keepaspectratio=true]{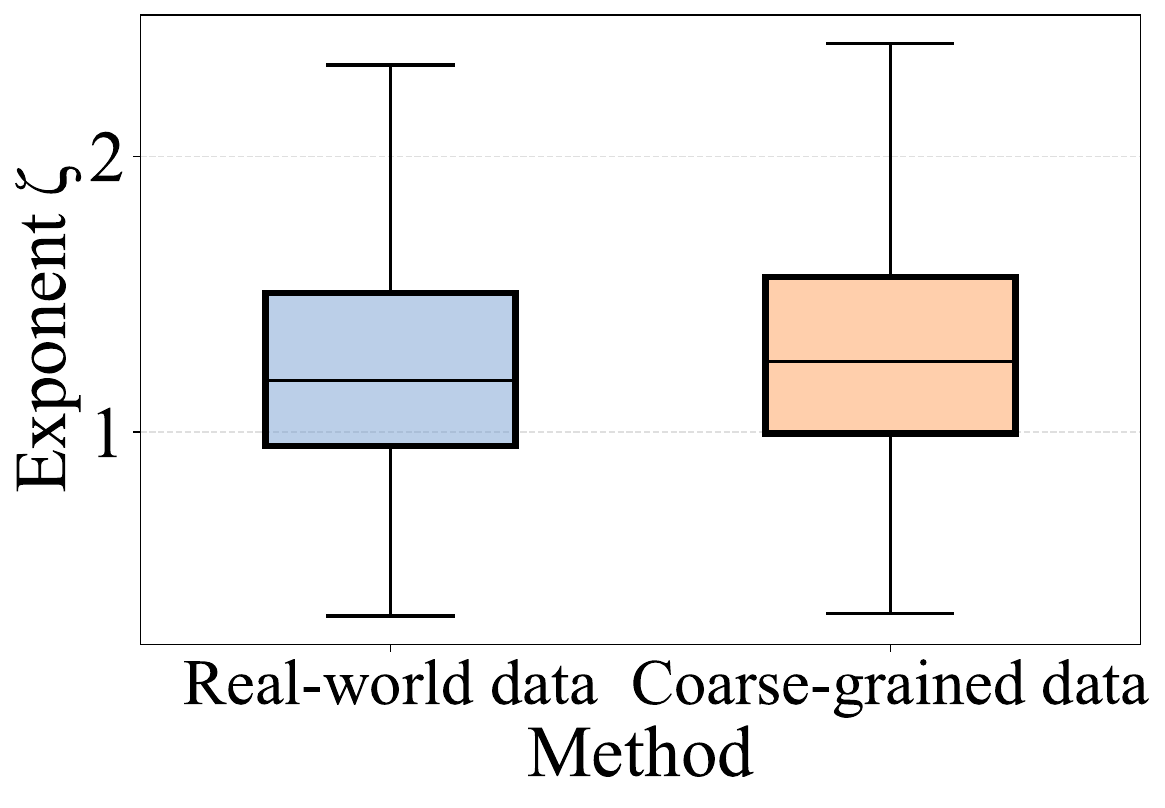}
    \vspace{-15pt}
    \captionsetup{font={small}}
    \caption{Visitation frequency (Real vs. Coarse-grained).}
    \label{fig:zipf_grid}
\end{minipage}
\end{figure}

In this section, we provide additional results that complement Section~\ref{sec:valid_coarse}.
Based on both real-world and coarse-grained data, we compare stay duration distributions and visitation frequency from temporal and spatial-temporal perspectives (Figures ~\ref{fig:duration_grid} and~\ref{fig:zipf_grid}).
We find that the distributions obtained from coarse-grained data closely match those obtained from the original real-world trajectories, indicating that coarse-graining preserves key scaling laws of mobility.

\subsection{\name{} optimization algorithm}

To make the optimization process easier to follow, we summarize the main steps of \name{} in Algorithm~\ref{alg:mcts}.

\begin{algorithm}[t]
\caption{\name{} optimization algorithm}
\label{alg:mcts}
\begin{algorithmic}[1]

\State \textbf{Input:} initial prompt state $s_0$, action space $\mathcal{A}$, target objectives $\mathbf{x}^*$, user sampling ratio $\rho$, number of iterations $N_{\mathrm{iter}}$
\State Initialize root node $v_0$ with state $s_0$

\For{$i=1$ to $N_{\mathrm{iter}}$}

    \State Initialize empty trajectory $\tau$

    \Comment{Selection}
    \While{$v$ has expanded children and $v$ is not terminal}
        \State $a \leftarrow \textsc{SelectActionByUCT}(v)$
        \State $\mathcal{U} \leftarrow \textsc{SelectUsersByUCB}(v.state,a,\rho)$
        \State Append $(v.state,a,\mathcal{U})$ to $\tau$
        \State $v \leftarrow \mathrm{child}(v,a)$
    \EndWhile

    \Comment{Expansion}
    \If{$v$ is not terminal}
        \State $\mathcal{C} \leftarrow \textsc{GlobalFilter}(v.state,\mathcal{A},B)$
        \ForAll{$a\in\mathcal{C}$}
            \State $\mathcal{U}_a \leftarrow \textsc{SelectUsersByUCB}(v.state,a,\rho)$
            \State $(r_a,s_a) \leftarrow \textsc{ApplyAndEvaluate}(v.state,a,\mathcal{U}_a,\mathbf{x}^*)$
        \EndFor
        \State $a^* \leftarrow \arg\max_{a\in\mathcal{C}} r_a$
        \State Create child node $v'$ with state $s_{a^*}$
        \State Add edge $(v,a^*)$ to the search tree
        \State Append $(v.state,a^*,\mathcal{U}_{a^*},r_{a^*})$ to $\tau$
        \State $v \leftarrow v'$
    \EndIf

    \Comment{Simulation}
    \State $s_{\mathrm{roll}} \leftarrow v.state$
    \For{$t=1$ to $L$}
        \If{$s_{\mathrm{roll}}$ is terminal}
            \State \textbf{break}
        \EndIf
        \State $\mathcal{C} \leftarrow \textsc{GlobalFilter}(s_{\mathrm{roll}},\mathcal{A},B)$
        \ForAll{$a\in\mathcal{C}$}
            \State $\mathcal{U}_a \leftarrow \textsc{SelectUsersByUCB}(s_{\mathrm{roll}},a,\rho)$
            \State $(r_a,s_a) \leftarrow \textsc{ApplyAndEvaluate}(s_{\mathrm{roll}},a,\mathcal{U}_a,\mathbf{x}^*)$
        \EndFor
        \State $a^* \leftarrow \arg\max_{a\in\mathcal{C}} r_a$
        \State Append $(s_{\mathrm{roll}},a^*,\mathcal{U}_{a^*},r_{a^*})$ to $\tau$
        \State $s_{\mathrm{roll}} \leftarrow s_{a^*}$
    \EndFor

    \Comment{Back-propagation}
    \State Update node-level statistics with cumulative returns
    \State Update global action statistics with immediate rewards
    \State Update user-level statistics for selected users

\EndFor

\State \Return the optimized prompt state from the search tree

\end{algorithmic}
\end{algorithm}

\subsection{Example of prompt adjustment}
To illustrate how behavioral constraints are injected into user profiles, we present two examples in which an initial profile is augmented with an additional behavioral description. The original profile contains only basic attributes, such as age, occupation, education level, consumption level, and coarse home/work locations. The adjusted profile further specifies mobility preferences that guide the generation model toward a particular behavioral pattern. For conciseness, the examples below report only the profile field. All other parts of the prompt, including the task description, formatting instruction remain fixed.
\begin{tcolorbox}[
colback=gray!2,
colframe=black!60,
    title={Example 1: prompt adjustment},
    fonttitle=\bfseries,
    arc=1mm,
    boxrule=0.5pt
]
\textbf{Initial profile:}
A 35-year-old male logistics worker in Beijing with a high-school education level and medium consumption level. His home is near $[116.5, 39.8]$, and his workplace is near $[116.7, 38.0]$.

\vspace{0.5em}
\textbf{Profile after adjustment:}
\textbf{[Initial profile]}.
\textcolor{blue!70!black}{
In addition, the user has very weak location preference and high variability in visited places.
His movements are largely driven by tasks, assignments, or service-related needs rather than routine.
There is no single dominant activity center, and his mobility naturally spans multiple areas.
}
\end{tcolorbox}

\begin{tcolorbox}[
colback=gray!2,
colframe=black!60,
    title={Example 2: prompt adjustment},
    fonttitle=\bfseries,
    arc=1mm,
    boxrule=0.5pt
]
\textbf{Initial profile:}
A 23-year-old female IT engineer in Beijing with a bachelor's degree and medium consumption level.
Her home is near $[116.3, 40.0]$, and her workplace is near $[116.3, 40.0]$.

\vspace{0.6em}
\textbf{Profile after adjustment:}
\textbf{[Initial profile]}.
\textcolor{blue!70!black}{
In addition, most of her activities are expected to occur near home or within the same neighborhood.
She has a strong nearest-option preference when choosing destinations, and tends to visit fixed local places repeatedly.
}
\end{tcolorbox}

\subsection{Prompt example}
\label{app:prompt}
We provide two representative prompt examples: one for the mobility simulation model and another for action generation in the MDP-based prompt adjustment process.
\begin{tcolorbox}[width=\linewidth, 
title=Abridged mobility simulation prompt example, 
breakable,
colback=gray!2,
colframe=black!60,
    arc=1mm,
    boxrule=0.5pt,
fonttitle=\bfseries]
\textbf{INPUT DATA}\par
\texttt{Profile: \{profile, today\_date, home\_location, work\_location\}.}\par
\texttt{Date: \{today\_date\}.}

\textbf{ROLE \& TASK}\par
You are a human mobility simulation agent. Given a person's profile, date, home location, work location, and candidate POIs, generate the person's daily mobility trajectory. The output should be a sequence of mobility events.

\textbf{TASK REQUIREMENTS}\par
\begin{itemize}
    \item Generate events in chronological order.
    \item Each event should correspond to a physical visit to a location with a clear purpose.
    \item Do \textbf{not} generate passive or non-mobility activities, such as sleeping, resting at home, watching TV, or browsing the phone.
    \item Do \textbf{not} generate continuous GPS traces; each output item should be an event-level spatiotemporal point.
    \item ...
\end{itemize}

\textbf{OUTPUT FORMAT}\par
1. At 8:12 a.m., commute from home to the logistics warehouse for the morning work shift. Activity: Work. Location type: workplace. Location: [116.7, 38.0].
\end{tcolorbox}

\begin{tcolorbox}[
width=\linewidth,
title=Abridged action generation prompt example,
label=box:daily-prompt,
breakable,
colback=gray!2,
colframe=black!60,
    arc=1mm,
    boxrule=0.5pt,
fonttitle=\bfseries
]

\textbf{\large ROLE}

You are an expert in human mobility modeling, urban science, statistical physics of mobility, and LLM-based synthetic population simulation...

\vspace{0.5em}

\textbf{\large BACKGROUND}

 The simulated data is created by feeding human personas (profiles) into an LLM, which generates mobility trajectories.
The objective is to improve simulation realism by adjusting persona design or constraining behavioral rules...

\vspace{0.5em}

\textbf{\large INPUT DATA}

\vspace{0.5em}

\textbf{\large ANALYSIS GOALS}

\begin{itemize}
\item Population composition biases
\item Missing or exaggerated activity-space archetypes
\item Behavioral causes of distribution mismatch
\item Ways to adjust personas or behavioral rules
\end{itemize}

\vspace{0.5em}

\textbf{\large TASK REQUIREMENTS}

\textbf{TASK 1 — Population-level Diagnosis}

From a human mobility perspective, briefly explain what the observed differences suggest about overall activity space. Describe whether the simulated population tends to move within larger or smaller areas than real people, and identify which general mobility archetypes may be overrepresented or underrepresented...

\vspace{0.5em}

\textbf{TASK 2 — Population Group}

Goal:Define a practical radius-based segmentation schema.

\textit{Requirements:}

\begin{itemize}
\item Define 4--6 radius groups that partition individuals by activity-space size (small $\rightarrow$ large).
\item Groups must be defined by radius magnitude.
\item For each group provide: Group name, Radius range (clear thresholds or relative scale), One-sentence behavioral description and Target proportion (\%) representing a realistic population distribution...
% \item ...
\end{itemize}

\vspace{0.5em}

\textbf{TASK 3 — Adjustment Strategy}

For each group defined in Task 2, provide concrete adjustments aimed at improving simulation realism...

Examples include:

\begin{itemize}
\item Adding behavioral constraints (e.g., stronger home/work anchors)
\item Increasing routine or location stability...
\end{itemize}

\vspace{0.5em}

\textbf{\large OUTPUT FORMAT}

\begin{enumerate}
\item Population Diagnosis
\item Population Groups Table

Columns:
Group Name $\mid$ Radius Range $\mid$ Behavioral Description $\mid$ Target  proportion %

\item Adjustment Strategy (group-by-group suggestions)
\end{enumerate}

\end{tcolorbox}

\end{document}